\documentstyle[12pt,epsf,epsfig,a4]{article}
\begin{document}
\begin {titlepage}
\begin{flushleft}
FSUJ TPI QO-8/97
\end{flushleft}
\begin{flushright}
August, 1997
\end{flushright}
\vspace{20mm}
\begin{center}
{\Large {\bf
Homodyne measurement of exponential phase moments}
\\[3ex]
\large M. Dakna, T. Opatrn\'{y}$^{\ast}$, D--G. Welsch}
\\[2.ex]
Friedrich-Schiller-Universit\"at Jena 
Theoretisch-Physikalisches Institut \\[1ex]
Max-Wien Platz 1, D-07743 Jena, Germany
\vspace{25mm}
\end{center}
\begin{center}
\bf{Abstract}
\end{center}
It is shown that the exponential moments of the canonical phase 
can be directly sampled from the data recorded in balanced 
homodyne detection. Analytical expressions for the sampling 
functions are derived, which are valid for arbitrary states and 
bridge the gap between quantum and classical phase. The reconstruction 
of the canonical phase distribution from the experimentally determined 
exponential moments is discussed.
\end{titlepage}
\renewcommand {\thepage} {\arabic{page}}
\setcounter {page} {2}
\section{Introduction}
\label{sec1}
Since Dirac's attempt in 1927 to introduce amplitude and phase operators 
in quantum mechanics \cite{Dirac} a number of concepts have been developed 
with the aim to overcome the problems resulting from the non-existence of 
a Hermitian phase operator (for a review, see \cite{Lynch}). Recently
an attempt has been made to bridge the gap between two concepts which are 
based on essentially different approaches to the phase problem and widely 
used in quantum optics \cite{Leonhardt1}. In the first, the phase of a 
radiation-field mode is defined from the requirement that phase and 
photon number should be complementary quantities. This first-principle 
definition leads to the {\em canonical phase} (also called London phase), 
the associated phase states being the right-hand eigenstates of  
a one-sided unitary exponential phase operator \cite{London}. In the second,
phase quantities are defined from the output observed in phase-sensitive 
measurements, such as eight-port homodyne detection. It is well known that 
in such a scheme the $Q$ function [or, in the case of non-perfect detection, 
a smoothed $Q$ function, i.e., an $s$-parametrized phase-space function with 
$s$ $\!<$ $\!-1$] is measured \cite{Walker,Noh}. The {\em measured phase} 
distribution can then be obtained from radially integrating the  
(smoothed) $Q$ function. Whereas in the classical limit the measured 
phase coincides with the canonical phase, in the quantum regime the 
two phases significantly differ from each other in general, because of
the additional noise unavoidably connected with the $Q$ function. So,
from a study of the asymptotic behaviour of the measured and
canonical phase distributions in the semiclassical domain it can be 
anticipated that the measured distribution is at least broader 
than the canonical one \cite{Leonhardt1}. In the quantum regime  
it is principally not possible to obtain the canonical phase distribution 
from the radially integrated $Q$ function, but it must be related to the 
complete quantum state, i.e., the complete $Q$ function as a
representation of the state in the phase space.    
     
The best and perhaps ultimate method for measuring the quantum state
of a traveling optical field has been four-port homodyne detection
in which the quantum state is measured in terms of the
quadrature-component distribution \cite{KVogel1}. Since the
quadrature-component distribution contains all knowable
information on the quantum state, the various quantum-statistical 
properties of the system can be obtained from it.
Moreover, the quadrature-component distribution is less noisy than the 
$Q$ function and therefore it is more suitable for determining the 
quantum statistics than the $Q$ function. The method also called {\em optical 
homodyne tomography} (OHT) was first used for reconstructing the Wigner 
function of a single-mode optical field applying inverse Radon transform 
\cite{Smithey1}, which requires a three-fold integration of the measured 
data. In the numerical calculation the standard filtered back projection 
algorithm is usually used, so that the reconstruction of the Wigner function 
is biased by data filtering. 

This problem does not appear and the effort is drastically reduced
if the quantities that one is interested in can be directly sampled from 
the homodyne data. In particular, the determination of the quantities 
and the error estimation are very fast and can be performed in {\em real
time}. Systematic errors can easily be reduced to any desired degree of 
accuracy and the remaining statistical errors only reflect the finite 
number of measurement events. It has been shown that both the density 
matrix in the photon-number basis \cite{Ariano1} and the moments and 
correlations of the photon creation and destruction operators 
\cite{Richter1} can be obtained in this way, which has offered novel 
possibilities of the experimental determination of the photon-number 
statistics of light.

In contrast to the photon-number statistics, the determination of
the canonical phase statistics has been an open problem. The phase
statistics can of course be tried to be determined indirectly by 
calculating it approximately from the Wigner function, using in the
reconstruction of the Wigner function the standard filtered back 
projection algorithm of inverse Radon transformation \cite{Smithey2}. 
Another indirect method, which avoids the rather lengthy detour via the 
Wigner function, is the calculation of the phase statistics from 
the sampled density matrix in the photon-number basis \cite{Schiller3}. 
However, since canonical phase and photon number are complementary quantities, 
there is no {\em a priori} upper bound for the density-matrix elements that 
can contribute to the phase statistics. Hence, large numbers of 
density-matrix elements must be sampled, the statistical errors of which 
then give rise to an error accumulation in the phase statistics such that 
the inaccuracies eventually dominate the result (note that the 
statistical error of the off-diagonal density-matrix elements increases 
with the distance from the diagonal). To limit the effect of inaccuracies,
one must necessarily restrict the method to states of low 
photon numbers and appropriately truncate them. A way that remains
to overcome the problem is to directly sample the phase statistics
from the homodyne data. Unfortunately the canonical phase distribution 
cannot be related to the quadrature-component distribution in the
sense of a sampling formula because the corresponding integral kernel
does not exist. It has been therefore suggested to introduce the exact phase 
distribution as the limit of a convergent sequence of appropriately 
parametrized (smeared) distributions each of which can directly be 
sampled from the homodyne data \cite{Dakna1}. The exact phase distribution 
can then be obtained asymptotically to any degree of accuracy, if the 
sequence parameter is chosen such that smearing is suitably weak. In 
practice it is therefore required that the sampling procedure is performed
simultaneously for various values of the sequence parameter, each 
value giving rise to its own sampling function. The disadvantage of the method 
is rather technical, since the numerical effort 
drastically increases with the number of photons contained in a state. This 
fact makes the method effectively applies only to states with low photon numbers.
Finally, it has been suggested to measure the canonical phase distribution 
asymptotically by replacing the local-oscillator in the homodyne
detection scheme with a reference mode prepared in so-called reciprocal 
binomial states -- a method that is also state dependent and hardly 
realizable at present \cite{Barnett}. 

In this paper we show that the problem of direct determination of the 
canonical phase statistics from the homodyne data can be solved
when it is based on the exponential phase moments (i.e., the Fourier 
components of the phase distribution) and not on the phase distribution
itself. We show that the exponential phase moments can be directly sampled 
from the quadrature-component distribution, without making a detour via 
other quantities and without any assumptions and approximations with regard 
to the state. In particular, we derive analytical expression for the 
sampling functions and give a very simple procedure for the numerical 
calculation. Since the method is independent of the state, it applies to 
both quantum and classical fields and all fields in between in a 
unified way and bridges, through the universally valid sampling functions, 
the gap between quantum and classical phase. Needless to say that
for obtaining the full information on the phase statistics, all 
(non-vanishing) exponential phase moments must be determined. 
It is worth noting that already sampling of a few low-order moments 
provide us with interesting information \cite{Opatrny1}.

The paper is organized as follows. In Sec.~\ref{S2} the problem of 
direct sampling of the exponential phase moments of a classical 
oscillator from the quadrature-component distribution is studied. 
In Sec.~\ref{S3} the theory is extended to the canonical phase 
of a quantum oscillator. Measurement errors are studied 
in Sec.~\ref{S4}, and in Sec.~\ref{S5} numerical results of computer 
simulations of measurements for determining the canonical phase
statistics are presented. Lengthy mathematical derivations are 
given in appendices.


\section{Sampling of exponential phase moments -- classical case}
\label{S2}

In order to gain insight into the problem of phase measurement
by means of balanced homodyne detection,
let us first consider the situation in classical optics. 
Here we can assume a proper phase-space probability  
$W(q,p)\,dqdp$, which can be rewritten as, on introducing polar 
coordinates $q$ $\!=$ $\!r \cos \varphi$ and 
\mbox{$p$ $\!=$ $\!r \sin \varphi$}, 
\begin{eqnarray}
\label{c1}
W(q,p)\, dqdp = 
P(r,\varphi) \, dr d\varphi ,
\end{eqnarray}
where 
\begin{eqnarray}
\label{c2}
P(r,\varphi) = r\,W(r \cos \varphi,r \sin \varphi). 
\end{eqnarray}
The phase probability distribution $P(\varphi)$ is then defined by 
\begin{eqnarray}
\label{c3}
P(\varphi) = \int_{0}^{\infty}  dr \, P(r,\varphi),
\end{eqnarray}
and the exponential phase moments $\Psi_{k}$, which are given by the 
Fourier components of the phase probability distribution,  
\begin{eqnarray}
\label{c4a}
\Psi_k &=& \int_{2\pi} d\varphi \, e^{ik\varphi} P(\varphi), 
\end{eqnarray}
can be written as 
\begin{eqnarray}
\label{c4}
\Psi_k &=& \int_{2\pi}d\varphi \int_{0}^{\infty} dr \, 
e^{ik\varphi} P(r,\varphi).
\end{eqnarray}
In balanced homodyne detection the probability distributions
$p(x,\vartheta)$ for the field quadratures $x(\vartheta)$ $\!=$ 
$\!q \cos \vartheta$ $\! +$ $\!p \sin \vartheta$ are measured.
[Note that when the harmonic oscillator represents a moving
particle in a harmonic potential, then $\vartheta =\omega t$
is valid and $x(\omega t)$ is the time-dependent position of the 
particle.] The quadrature-component probability distribution $p(x,\vartheta)$ 
can be obtained from the phase-space probability distribution $P(r,\varphi)$
as, on recalling that $x(\vartheta)$ 
$\!=$ $\!r \cos(\varphi \!-\! \vartheta)$,
\begin{eqnarray}
\label{c5}
p(x,\vartheta) = \int_{2\pi} d\varphi \int_{0}^{\infty} dr \,
P(r,\varphi) \, \delta [x\!-\!r \cos(\varphi \!-\! \vartheta)] ,
\end{eqnarray}
which is nothing but the well-known Radon transform, whose inversion
yields the phase-space probability distribution in terms of the
quadrature-component distribution.

Let us now turn to the problem of direct sampling of the exponential
phase moments. 
A quantity ${\cal A}$ can be determined from the homodyne data by means of 
the sampling method, if it can be related to $p(x,\vartheta)$ as
\begin{eqnarray}
\label{c6a}
{\cal A} =\int_{2\pi} d \vartheta \int_{-\infty}^{\infty} dx
\, K_{\cal A}(x,\vartheta)\,p(x,\vartheta) 
\end{eqnarray}
with a well-behaved integral kernel $K_{\cal A}(x,\vartheta)$ as sampling 
function. Note that $p(x,\vartheta+\pi)$ $\!=$ $\!p(-x,\vartheta)$, so 
that the $\vartheta$ integration in Eq.~(\ref{c6}) can be restricted to a 
$\pi$ interval. For the sake of convenience, here and in the following 
we prefer a $2\pi$ interval. 
In contrast to the full phase-space probability distribution $P(r,\varphi)$, 
which cannot be obtained from $p(x,\vartheta)$ by a simple inversion of
Eq.~(\ref{c5}) in the form of Eq.~(\ref{c6a}), the sampling method 
applies to the Fourier components of the radially integrated phase-space 
probability distribution, i.e., the exponential phase moments $\Psi_{k}$
can be given by 
\begin{eqnarray}
\label{c6}
\Psi_k =\int_{2\pi} d \vartheta \int_{-\infty}^{\infty} dx
\, K_{k}(x,\vartheta)\,p(x,\vartheta) .
\end{eqnarray}
Substituting in Eq.~(\ref{c6}) for $p(x,\vartheta)$ the
expression (\ref{c5}), we arrive at
\begin{eqnarray}
\label{c7}
\Psi_k =
\int_{2\pi} d \vartheta \int_{2\pi} d \varphi 
\int_{-\infty}^{\infty} dx \int_{0}^{\infty} dr
\, K_{k}(x,\vartheta) 
\, P(r,\varphi) \, \delta [x\! -\! r \cos (\varphi\! - \!\vartheta)] .
\end{eqnarray}
In order to determine the integral kernel $K_{k}(x,\vartheta)$,  
it is convenient to introduce the Fourier decomposition
\begin{eqnarray}
\label{ce1}
K_{k}(x,\vartheta)
= \sum_{l=-\infty}^{\infty}e^{il\vartheta} 
K_{k,l}(x) .
\end{eqnarray}
When the phase argument $\varphi$ in $P(r,\varphi)$ is shifted
towards $\varphi$ $\!+$ $\!\varphi_{0}$, i.e., $P(r,\varphi)$ 
$\!\to$ $\!P(r,\varphi + \varphi_{0})$, then it follows
from Eq.~(\ref{c4}) that $\Psi_{k}$ changes as
$\Psi_{k}$ $\!\to$ $\!e^{-ik\varphi_{0}}\Psi_{k}$. 
Comparing this requirement with Eqs.~(\ref{c7}) and (\ref{ce1}),
we find that $K_{k,l}(x)$ must be of the form $K_{k,l}(x)$
$\!=$ $\!\delta_{k,l}K_{k}(x)$, and hence
\begin{eqnarray}
\label{c9}
K_{k}(x,\vartheta) = e^{ik\vartheta} K_{k}(x).
\end{eqnarray}
We now insert this expression into Eq.~(\ref{c7}), compare 
the result with Eq.~(\ref{c4}) and find that $K_{k}(x)$ must 
satisfy the integral equation
\begin{eqnarray}
\label{cl3}
\int_{2\pi} d\varphi \, e^{ik\varphi} K_{k}(r \cos \varphi) = 1
\end{eqnarray}
for all $r$ $\!>$ $\!0$. 

   From Eq.~(\ref{cl3}) we can see that $K_{k}(x)$ is not uniquely defined. 
First, any function of parity $(-1)^{k+1}$ can be added to $K_{k}(x)$ 
without changing the integral. Second, any polynomial of a degree less than 
$k$ can also be added to $K_{k}(x)$. As can be verified by direct 
substitution, a solution of Eq.~(\ref{cl3}) for odd and even
$k$, respectively, is given by
\begin{eqnarray}
\label{cl4}
K_{2m+1}(x) = {\textstyle\frac{1}{4}}
(-1)^{m} (2m+1)\, {\rm sign}\,(x)
\end{eqnarray}
and
\begin{eqnarray}
\label{cl5}
K_{2m}(x) = \pi^{-1} (-1)^{m+1} m \ln|x|+C
\end{eqnarray}
[$m$ $\!=$ $\!0,1,2,\ldots$,where  $C$ is an (irrelevant) constant.] Note that this solution ensures
that the integral (\ref{c6}) exists for any quadrature-component 
distribution $p(x,\vartheta)$ which with increasing $|x|$ decreases
at least as $|x|^{-(1+\epsilon)}$, $\epsilon$ being a (arbitrarily 
small) positive constant, i.e., for any physical state. Clearly, this would be 
not the case if, within the ambiguity mentioned, polynomials were added to the
functions (\ref{cl4}) and (\ref{cl5}). Another reason for choosing 
the functions (\ref{cl4}) and (\ref{cl5}) is the reduction of the 
statistical error in a real experiment. Since this error is related 
to the variance of the kernel (Sec.~\ref{Ser}), it is advantageous to 
choose kernels which are varying as slowly as possible.


\section{Sampling of exponential phase moments -- quantum case}
\label{S3}

It is worth noting that the results derived in Sec.~\ref{S2}  
also remain valid for a quantized radiation-field mode, provided that 
$W(q,p)$ [or in polar coordinates, $P(r,\varphi)$] is identified with the 
quantum-mechanical Wigner function. Hence, using in Eqs.~(\ref{c7})
and (\ref{c9}) the functions (\ref{cl4}) and (\ref{cl5}) enables one to 
determine exponential phase moments defined by the Fourier components of 
the radially integrated Wigner function from the homodyne data by means
of the sampling method. Since the Wigner function of a quantum 
oscillator can attain negative values, it cannot be regarded, in general, 
as a proper phase-space probability distribution, and hence the radially 
integrated Wigner function does not represent, in general, a proper phase 
distribution function.

As already mentioned, for a quantized radiation-field mode photon number 
and canonical phase are complementary variables, and in place of 
Eq.~(\ref{c3}) we have 
\begin{eqnarray}
\label{qic1}
P(\varphi) = (2\pi)^{-1}  \langle \varphi | \hat \varrho | \varphi
\rangle , 
\end{eqnarray}
where $\hat{\varrho}$ and $|\varphi \rangle$, respectively, are the density 
operator of the state and the (unnormalizable) phase states \cite{Susskind1}
\begin{eqnarray}
\label{qic2}
|\varphi \rangle = \sum_{n=0}^{\infty} e^{in\varphi} |n\rangle ,
\end{eqnarray}
which are right-hand eigenstates of the only one-sided unitary
operator
\begin{eqnarray}
\label{qic2a}
\hat E = (\hat n +1)^{1/2} \, \hat a,
\end{eqnarray}
\begin{eqnarray}
\label{qic2b}
\hat E |\varphi \rangle = e^{i\varphi} |\varphi \rangle.
\end{eqnarray}
In Eq.~(\ref{qic2a}), $\hat n$ $\!=$ $\!\hat a^{\dag} \hat a$ is 
the photon-number operator, $\hat a^{\dag}$ and $\hat a$ being the 
photon creation and annihilation operators, respectively. From 
Eqs.~(\ref{qic1}) and (\ref{qic2b}) together with the fact
that the phase states resolve the unity it is easily seen that
the exponential phase moments $\Psi_{k}$ defined in Eq.~(\ref{c4a})
can be written as
\begin{eqnarray}
\label{qic3}
\Psi_{k} = \langle \hat E^{k} \rangle
\end{eqnarray}
for $k$ $\!=$ $\!1,2,\dots$, and $\Psi_{k}$ $\!=$ $\!\Psi_{-k}^{*}$ 
for $k$ $\!=$ $\!-1,-2,\dots$. We now combine Eqs.~(\ref{qic2a}) and
(\ref{qic3}) and obtain [in place of Eq.~(\ref{c4})] 
\begin{eqnarray}
\label{qi1}
\Psi_k = \sum_{n=0}^{\infty}
\varrho_{n\!+\!k,n} 
\end{eqnarray}
($k$ $\!=$ $\!1,2,\dots$). Next, we express the 
quadrature-component distribution
\begin{eqnarray}
\label{qi2a} 
p(x,\vartheta) = \langle x,\vartheta | \hat{\varrho} | x,\vartheta \rangle
\end{eqnarray}
in terms of the density-matrix elements in the photon-number basis.
For this purpose we expand in Eq.~(\ref{qi2a}) the eigenstates 
$| x,\vartheta \rangle$ of the quadrature-component operator 
$\hat{x}(\vartheta)$ $\!=$ $\!2^{-1/2}$ $\!(e^{-i\vartheta}\hat{a}$
$\!+$ $\!e^{i\vartheta}\hat{a}^{\dagger})$
in the photon-number basis \cite{VoWe},
\begin{eqnarray}
\label{qi2c} 
| x,\vartheta \rangle =
\sum_{n=0}^{\infty} e^{in\vartheta} \psi_{n}(x) \, |n\rangle,
\end{eqnarray}
where the functions $\psi_{n}(x)$ are the eigenfunctions of the 
harmonic-oscillator Hamiltonian, $\psi_{n}(x)$ $\!=$ 
$\!(2^{n}n!\sqrt{\pi})^{-1/2}$ exp$(-x^{2}/2){\rm H}_{n}(x)$, 
H$_{n}(x)$ being the Hermite polynomial. From Eqs.~(\ref{qi2a})
and (\ref{qi2c}) we then obtain [in place of Eq.~(\ref{c5})]
\begin{eqnarray}
\label{qi3}
p(x,\vartheta) = \sum_{n\!=\!0}^{\infty} \sum_{m\!=\!0}^{\infty}
\psi_{n}(x) \psi_{m}(x)
\varrho_{m,n} e^{i(n-m)\vartheta} .
\end{eqnarray}

Let us again assume that $\Psi_{k}$ can be obtained from $p(x,\vartheta)$
according to Eq.~(\ref{c6}). Substituting in Eq.~(\ref{c6}) for 
$p(x,\vartheta)$ the quantum-mechanical expression (\ref{qi3}) and 
comparing the result with Eq.~(\ref{qi1}), we see that the kernel 
$K_{k}(x,\vartheta)$ must be of the form (\ref{c9}), 
but now $K_{k}(x)$ must satisfy the integral equation
\begin{eqnarray}
\label{qi5}
2\pi\int_{-\infty}^{\infty} dx \,
K_{k}(x) \psi_{n\!+\!k}(x) \psi_{n}(x) = 1
\end{eqnarray}
($n$ $\!=$ $\!0,1,2,\dots$). Equation~(\ref{qi5}) plays the same role for 
a quantum oscillator as Eq.~(\ref{cl3}) for a classical oscillator does. 
   From Eq.~(\ref{qi5}) and the properties of the Hermite polynomials the 
same ambiguity in the determination of $K_{k}(x)$ as in the classical 
case (Sec.~\ref{S2}) is found. Provided that a $K_{k}(x)$ exists,
any function of parity $(-1)^{k+1}$ and/or any polynomial of a degree 
less than $k$ can be added to $K_{k}(x)$ in order to again obtain a 
solution of Eq.~(\ref{qi5}).

We now turn to the problem of construction of an integral kernel that 
satisfies Eq.~(\ref{qi5}). For this purpose we return to 
Eq.~(\ref{qic3}) and bring the operator $\hat E^{k}$ into the normally 
ordered form,
\begin{eqnarray}
\hat E^{k} & = & \sum_{n=0}^\infty
:\frac{\hat{a}^{\dagger n} \exp(-\hat{a}^\dagger \hat{a})
\hat{a}^{n+k}} {\sqrt{n!(n+k)!}}:
\nonumber \\ & = & 
\sum_{n=0}^\infty\sum_{m=0}^\infty
\frac{1}{\sqrt{n!(n+k)!}}\frac{(-1)^m}{m!}
\hat{a}^{\dagger n+m}\hat{a}^{n+m+k}
\label{Ma3}
\end{eqnarray}
(the notation $:\ :$ is used to indicate normal ordering). From the 
expansion (\ref{Ma3}) together with the sampling formula for normally 
ordered moments and correlations of the photon creation and 
annihilation operators \cite{Richter1},
\begin{eqnarray}
\langle \hat{a}^{\dagger n}\hat{a}^m\rangle
= \left[2\pi \sqrt{2^{n+m}} {n\!+\!m\choose m}\right]^{-1}
\!\!\int_{2\pi} \! d \vartheta \!\! \int_{-\infty}^{\infty} \! dx \, 
e^{i(n-m)\vartheta} \, {\rm H}_{n+m}(x)
p(x,\vartheta),
\label{Ma1}
\end{eqnarray}
we find after some calculation (see \ref{app0}) that $\Psi_{k}$, 
Eq.~(\ref{qic3}), can be written in the form of Eq.~(\ref{c6}), 
\begin{eqnarray}
\label{Ma1xa}
\Psi_{k} = \int_{2\pi} d\vartheta \int_{-\infty}^{\infty} dx 
\, \tilde K_{k}(x) e^{ik\vartheta} p(x,\vartheta) .
\end{eqnarray}
In Eq.~(\ref{Ma1xa}), the integral kernel $\tilde K_{k}(x)$ can be 
decomposed into two parts,
\begin{eqnarray}
\label{kt1}
\tilde K_{k}(x) = K_{k}(x) - {\rm F}_{k}(x),
\end{eqnarray}
where for odd and even $k$, respectively, $K_{k}(x)$ reads as
\begin{eqnarray}
\lefteqn{
K_{2m+1}(x) = (-1)^m \, \frac{ 2x (m\!+\!1)! }{ (2\pi)^{m+3/2} }
\int_{0}^{+\infty} dr \, 
\Bigg\{
\Omega^{(2m+1)}(r^2)
}
\nonumber \\ && \hspace{15ex} \times \,
\frac{ r^{2m}  
\Phi[m\!+\!2,3/2,-x^2\tanh(r^2/2)]}
{e^{-(m+1)r^2} \sinh^m(r^2/2) \cosh^{m+2}(r^2/2)}
\Bigg\} 
\label{A1}
\end{eqnarray}  
and
\begin{eqnarray}
\lefteqn{
K_{2m}(x) = (-1)^m \, \frac{m!}{(2\pi)^{m+1}}
\int_{0}^{+\infty} dr \,
\Bigg\{
\Omega^{(2m)}(r^2)
}
\nonumber \\ && \hspace{10ex} \times \,
\frac{ r^{2m-1} e^{m r^2 / 2} }
{ \sinh^m(r^2/2) }
\bigg[
\frac{ \Phi[m\!+\!1, 1/2,- x^2 \tanh(r^2/2)] }
{ e^{-(m+1)r^2/2} \cosh^{m+1}(r^2/2)}-1
\bigg]
\Bigg\} ,
\label{A2}
\end{eqnarray}
and ${\rm F}_k(x)$ is the polynomial 
\\ \parbox{\textwidth}{
\begin{eqnarray}
\lefteqn{
{\rm F}_k(x) = 
\frac{1}{2\pi 2^{k/2}}
\sum_{n=1}^{[\frac{k-1}{2}]}
\Bigg[
\frac{(-2)^n(k-n)!}{(k-2n)!} \, {\rm H}_{k-2n}(x)
}
\nonumber \\ && \hspace{15ex} \times\,
\sum_{l=0}^{\infty}{n\!+\!l\!-1\choose l}
\frac{1}{\sqrt{(l+1)\dots(l+k)}} 
\Bigg] .
\label{Ma10}
\end{eqnarray} 
}
In Eqs.~(\ref{A1}) and (\ref{A2}), $\Phi (a,b,y)$ is the confluent 
hypergeometric function and $\Omega^{(k)}(z)$ defined 
in Eq.~(\ref{defo}) [together with Eq.~(\ref{sphercoord})] in 
\ref{app0} can be given by power-series expansion (\ref{app1}),
\begin{eqnarray}
\label{omega1}
\Omega^{(k)}(z) = \sum_{m=0}^{\infty} A_{m}^{(k)} \, z^{m} ,
\end{eqnarray}
where
\begin{eqnarray}
\label{omega2}
A_{m}^{(k)} =\frac{(-1)^m}{m!} 
\frac{2\pi^{k/2}}{\Gamma (k/2+m)} 
\left.
\frac{\partial ^{m}}{\partial x^{m}}
\left[ \prod_{j=1}^{k} (1-jx)^{-\frac{1}{2}} \right] 
\right|_{x=0} .
\end{eqnarray}
  
\begin{figure}
\centering\epsfig{figure=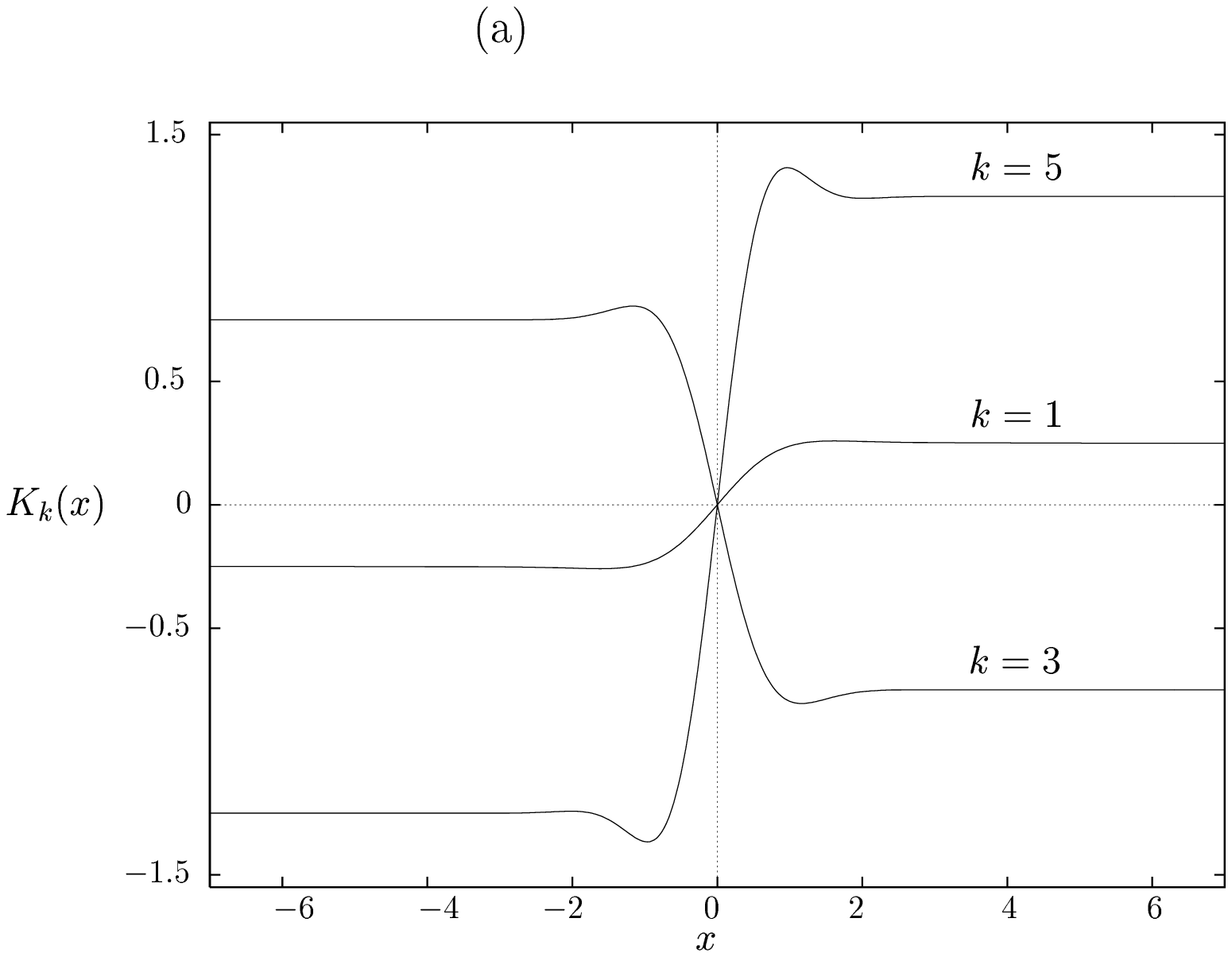,width=0.7\linewidth}

~
\vspace{0.1cm}

~

\centering\epsfig{figure=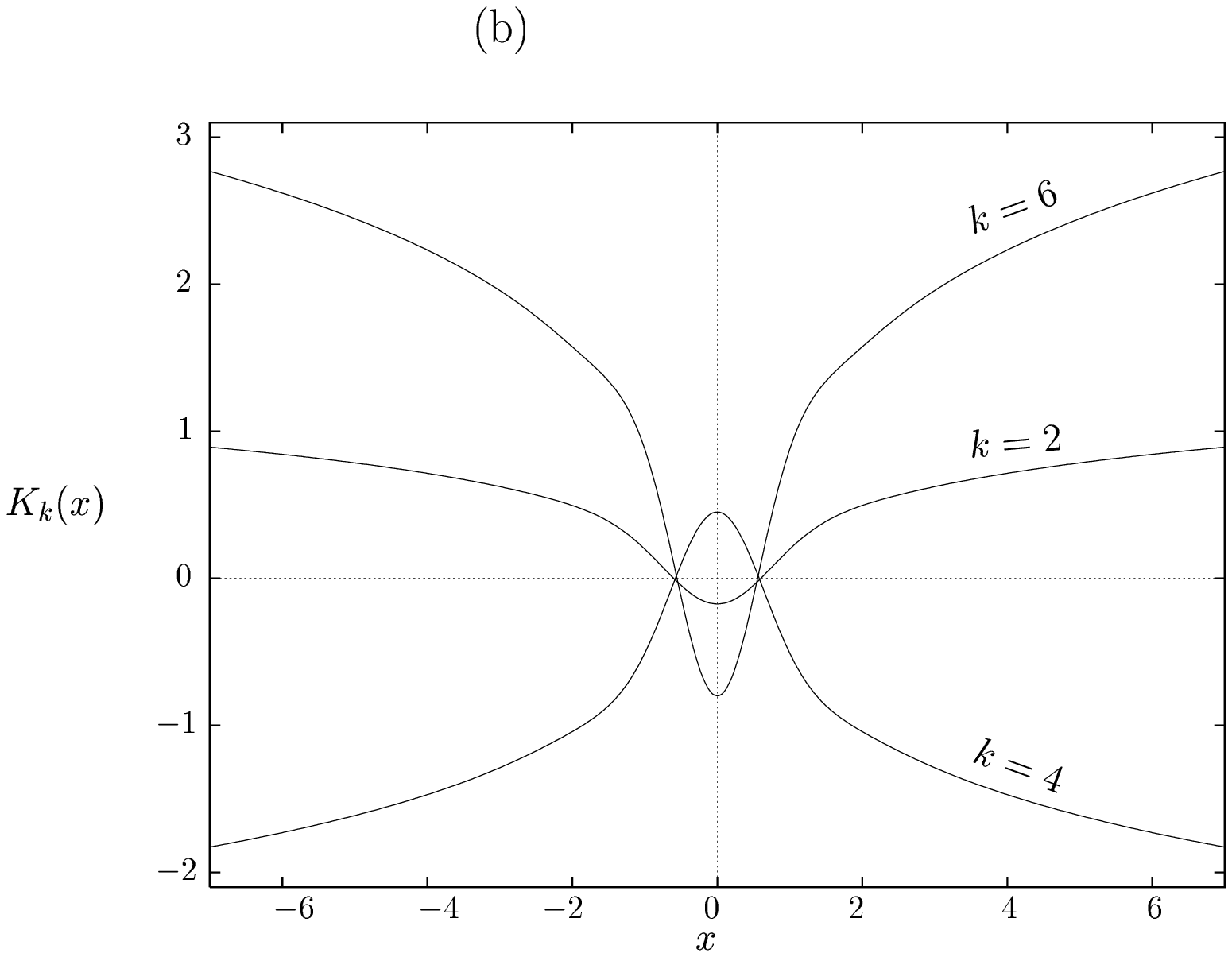,width=0.7\linewidth}
\caption{The $x$-dependent part $K_{k}(x)$ of the sampling function 
$K_{k}(x,\vartheta)$ $\!=$ $\!e^{ik\vartheta} K_{k}(x)$ for the 
determination of the exponential phase moments $\Psi_{k}$ from the 
quadrature-component distribution $p(x,\vartheta)$ is shown for various
odd (a) and even (b) $k$.
\label{F1}}
\end{figure}

  From Eqs.~(\ref{A1}) -- (\ref{Ma10}) [together with 
Eq.~(\ref{defo})] it is seen that $\tilde K_{k}(x)$ 
exists, and it can be proved by direct substitution that $\tilde K_{k}(x)$ 
satisfies Eq.~(\ref{qi5}). Hence we have found a solution of Eq.~(\ref{qi5})
even when the assumption made for constructing it fails (i.e., when
the moments and correlations (\ref{Ma1}) do not exist for all $m$ and
$n$, see the derivation in \ref{app0}). It is worth noting that both 
$\tilde K_{k}(x)$ and $K_{k}(x)$ are solutions of Eq.~(\ref{qi5}), 
because the polynomial F$_{k}(x)$ in Eq.~(\ref{kt1}) reflects the 
above mentioned ambiguity in the solution of Eq.~(\ref{qi5}) and 
can therefore be omitted. Further, it can be shown (\ref{app2})
that with increasing $|x|$ the solution $K_{k}(x)$  approaches 
the classical one, i.e., the asymptotic behavior for large $|x|$ of 
Eqs.~(\ref{A1}) and (\ref{A2}), respectively, is exactly given by 
Eq.~(\ref{cl4}) and
Eq.~(\ref{cl5}). 
We see that $K_{k}(x)$ can be used for determining exponential
phase moments from the homodyne data for all states whose 
quadrature-component distributions $p(x,\vartheta)$ asymptotically 
decrease at least as $|x|^{-(1+\epsilon)}$, $\epsilon$ $\!>$ $\!0$,
i.e., for any physical state. 
Since $K_{k}(x)$ applies to both quantum and classical systems in a 
unified way, the sampling method bridges the gap between quantum and 
classical phase. Examples of $K_{k}(x)$ for various $k$ are shown in 
Fig.~\ref{F1}. It is seen that they are well-behaved functions, which rapidly 
approach the classical limit and differ from it only in the small region 
of vacuum fluctuations.

Let us comment on the numerical calculation of $K_{k}(x)$ which can
be performed in a straightforward manner. In particular, from
Eqs.~(\ref{A1}) and (\ref{A2}), respectively, it can be easily proved that 
\cite{Opatrny1,DAriano}
\begin{eqnarray}
\label{ker1}
K_{1}(x) = \pi^{-3/2} 
x \int_{0}^{\infty} \frac{dt}{\sqrt{t}\,{\rm cosh}^{2}t}
\, \Phi\!\left( 2, {\textstyle\frac{3}{2}}, -x^2 {\rm tanh}\,t \right)
\end{eqnarray}
and [after calculating $\Omega^{(2)}(r^{2})$ according to Eq.~(\ref{defo})]
\cite{Opatrny1}
\begin{eqnarray}
\label{ker2}
K_{2}(x) = \frac{1}{2\pi}
\int_{0}^{\infty} dt \, {\rm I}_{0}(t) 
\Bigg[
\frac{e^{-2t}}{{\rm sinh}\ t} 
 \, - \, \frac{1}{{\rm cosh}^{2} t \, {\rm sinh}\,t} \,
\Phi\!\left( 2, \textstyle\frac{1}{2}, -x^{2} {\rm tanh}\,t \right) 
\Bigg] 
\end{eqnarray}
[I$_{0}(t)$, modified Bessel function]. 
The one-dimensional integrals can then be calculated numerically using
standard methods. In order to calculate $K_{k}(x)$ for arbitrary $k$ 
it may be convenient to start from
$\tilde K_{k}(x)$ as given in \mbox{Eq.~(\ref{05})} in \ref{app0} 
and approximate it in a small (nonclassical) interval by a finite 
sum such that 
\begin{eqnarray}
K_{k}(x) \approx (2\pi)^{-1}\sum_{l=0}^{l_{0}}
C_l^{(k)}{\rm H}_{2l+k}(x)+{\rm F}_k(x), 
\quad |x| < x_{0},
\end{eqnarray}
and use the classical limit, Eqs.~(\ref{cl4}) and (\ref{cl5}), elsewhere.
For example, for the calculation of the kernels plotted in Fig.~\ref{F1} 
it is sufficient to chose the parameters $l_{0}$ $\!=$ $\!40$ and 
$x_{0}$ $\!=$ $\!4$, and to truncate the infinite sum in Eq.~(\ref{Ma10}) 
for F$_{k}(x)$ at $l$ $\!=$ $\!10^{3}$.


\section{Nonperfect detection and measurement errors}
\label{S4} 
 
In practice there is always a number of experimental
inaccuracies that limit the precision with which  the exponential
phase moments can be determined. In this section we restrict attention
to three kinds of inaccuracies: data smearing, 
discretization of the phase parameter $\vartheta$, 
and finite number of measurement events (i.e., discretization of $x$).


\subsection{Data smearing}
\label{Losses}

Since in a realistic experiment the quadrature components cannot  
be measured with infinite precision, we may assume that instead of 
$p(x,\vartheta)$ a smeared distribution 
\begin{eqnarray}
\label{smear1}
p(x,\vartheta;\eta) = 
\int_{-\infty}^{\infty} dy \, f(x-y;\eta) \, p(y,\vartheta)  
\end{eqnarray}
is obtained. In Eq.~(\ref{smear1}), $f(x;\eta)$ is some positive 
single-peaked function and $\eta$ is a parameter quantifying the 
smearing effect. A typical example of $f(x;\eta)$ is a Gaussian, such as
\begin{eqnarray}
\label{smear2}
f(x;\eta) =  \frac{1}{\sqrt{2\pi\sigma^2}} 
\exp \!\left( - \frac{x^2}{2\sigma^2}\right),
\quad \sigma^2 = \frac{1-\eta}{2\eta} \, ,
\end{eqnarray}
which corresponds to the use of nonperfect photodetectors whose 
efficiency $\eta$ is less than unity. 

\begin{figure}
\centering\epsfig{figure=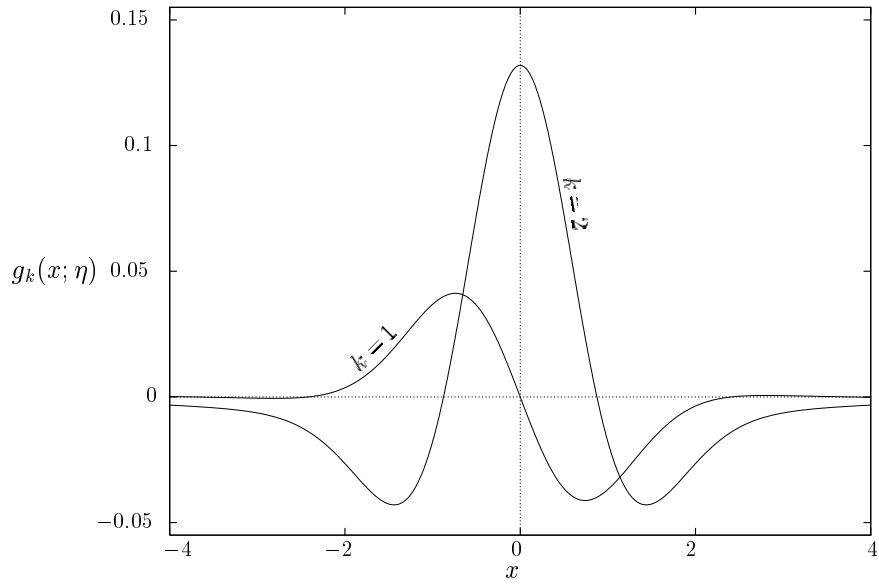,width=0.8\linewidth}
\caption{The function $g_{k}(x;\eta)$ for the determination of the systematic 
error $\Delta^{\rm (s)}\Psi_{k}$, Eq.~(\protect\ref{smear3a}),
which is associated with Gaussian data smearing, is shown for 
$k$ $\!=$ $\!1,2$ and $\eta$ $\!=$ $\!0.6$.
\label{F2}}
\end{figure}

Substituting in Eq.~(\ref{c6}) for the exact distribution $p(x,\vartheta)$
the smeared distribution $p(x,\vartheta;\eta)$ yields 
exponential phase moments $\Psi_{k}(\eta)$ that differ from $\Psi_{k}$ 
in a systematic error (bias) $\Delta^{\rm (s)}\Psi_{k}$ as follows:
\begin{eqnarray}
\label{smear3}
\Psi_{k}(\eta) = \int_{2\pi} d\vartheta 
\int_{-\infty}^{\infty} dx \,
K_{k}(x,\vartheta) \, p(x,\vartheta;\eta)
= \Psi_{k} + \Delta^{\rm (s)}\Psi_{k}
\end{eqnarray}
with
\begin{eqnarray}
\label{smear3a}
\Delta^{\rm (s)}\Psi_{k} =
\int_{2\pi} d\vartheta \, e^{ik\vartheta}
\int_{-\infty}^{\infty} dx \, g_{k}(x;\eta) \, p(x,\vartheta), 
\end{eqnarray}
where
\begin{eqnarray}
\label{smear4}
g_{k}(x;\eta) = \int_{-\infty}^{\infty} dy \, 
K_{k}(y) \left[f(y\!-\!x;\eta) - \delta (y\!-\!x) \right] .
\end{eqnarray}
Examples of the kernel $g_{k}(x;\eta)$ for the determination of the 
systematic error $\Delta^{\rm (s)}\Psi_{k}$ are plotted in Fig.~\ref{F2} 
[with $f(x;\eta)$ according to Eq.~(\ref{smear2})]. From a 
comparison of $g_{k}(x;\eta)$ with $K_{k}(x)$ (see Figs.~\ref{F1} and 
\ref{F2}) it is expected that the absolute values of $\Psi_{k}(\eta)$ are 
smaller than those of $\Psi_{k}$ in general.
The systematic error is state-dependent as it can be seen from
Eq.~(\ref{smear3a}). To give an impression of its magnitude, let us
restrict attention to the classical limit and consider a state whose
phase-space probability distribution is radially sharply localized    
at $r$ $\!=$ $\!r_{0}$ such that $r_{0}$ $\!\gg$ $\!\sigma$. In this
case it can be shown that, on assuming Gaussian smearing and using 
the results in Sec.~\ref{S2},    
\begin{eqnarray}
\label{smear5}
\Psi_{k}(\eta) 
\approx \exp\!\left( - \frac{k^{2}\sigma ^{2}}{2 r_{0}^{2}} 
\right) \Psi_{k} ,
\end{eqnarray} 
which reveals that the exponential phase moments can be determined from 
the smeared data quite reliably as long as $k$ $\!\ll$ $\!r_{0}/\sigma$.

\begin{figure}
\noindent
\begin{minipage}[b]{0.45\linewidth}
\centering\epsfig{figure=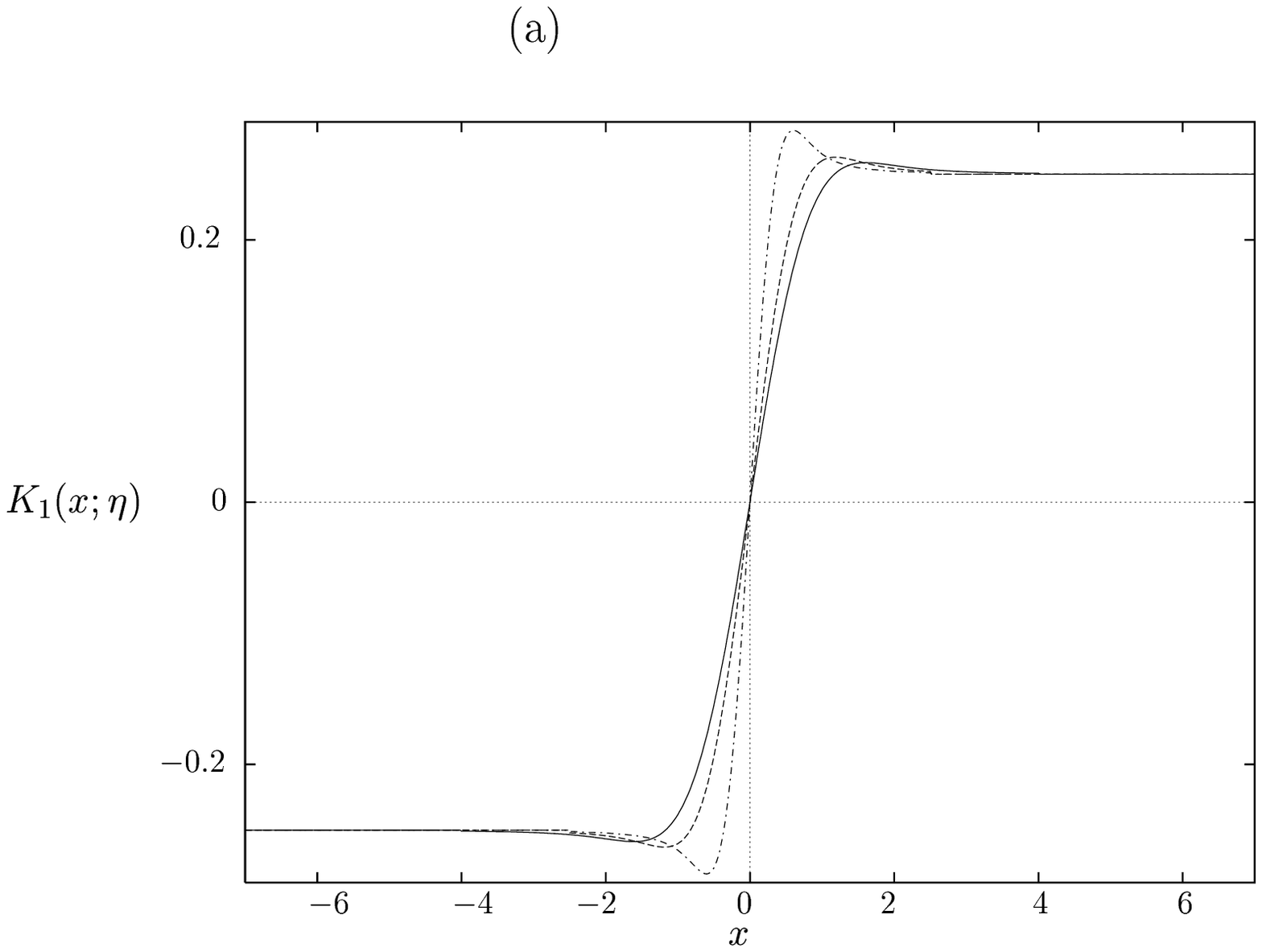,width=\linewidth}
\end{minipage}\hfill
\begin{minipage}[b]{0.45\linewidth}
\centering\epsfig{figure=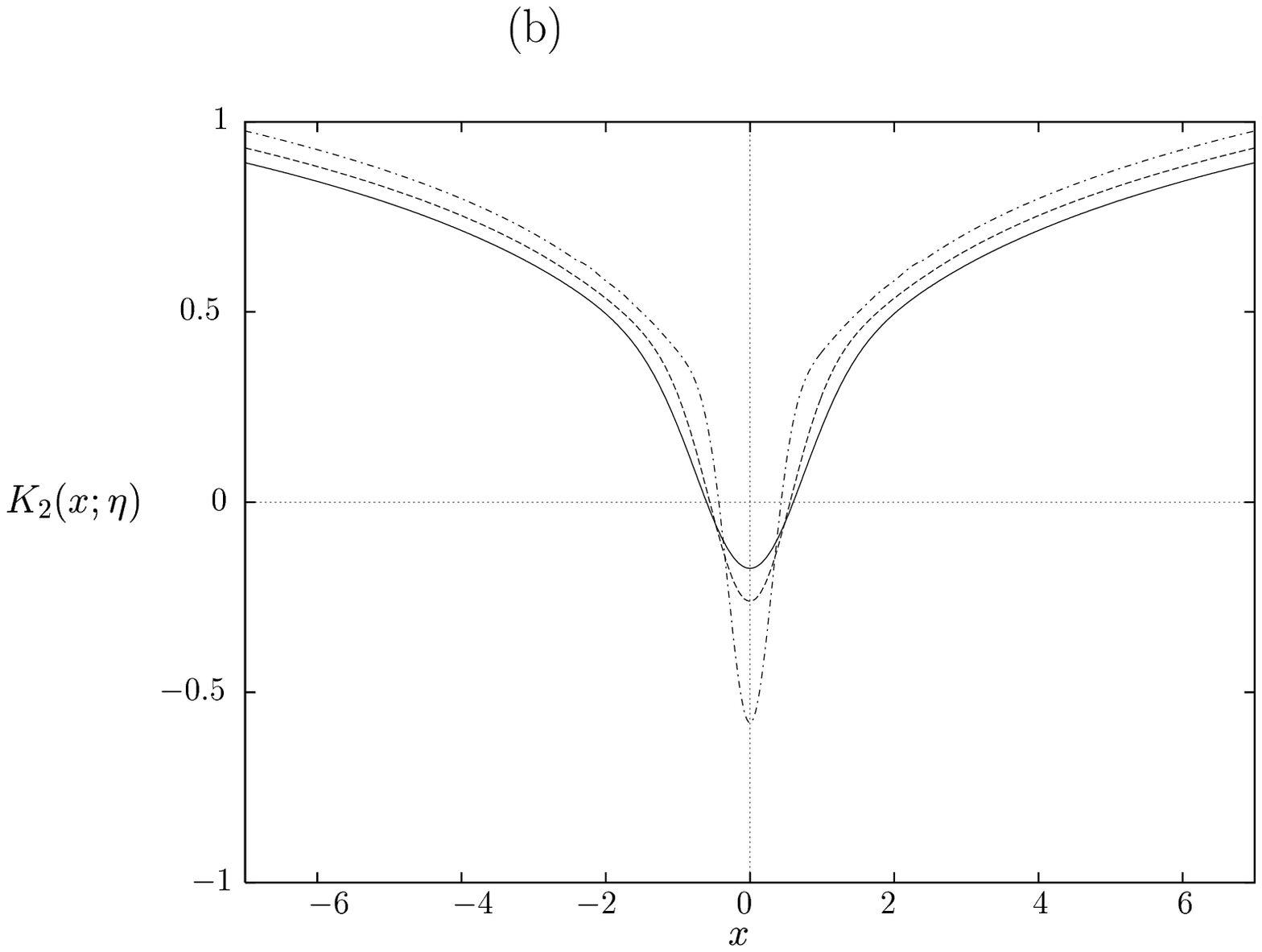,width=\linewidth}
\end{minipage}

~

\vspace{0.5cm}

~

\begin{minipage}[b]{0.45\linewidth}
\centering\epsfig{figure=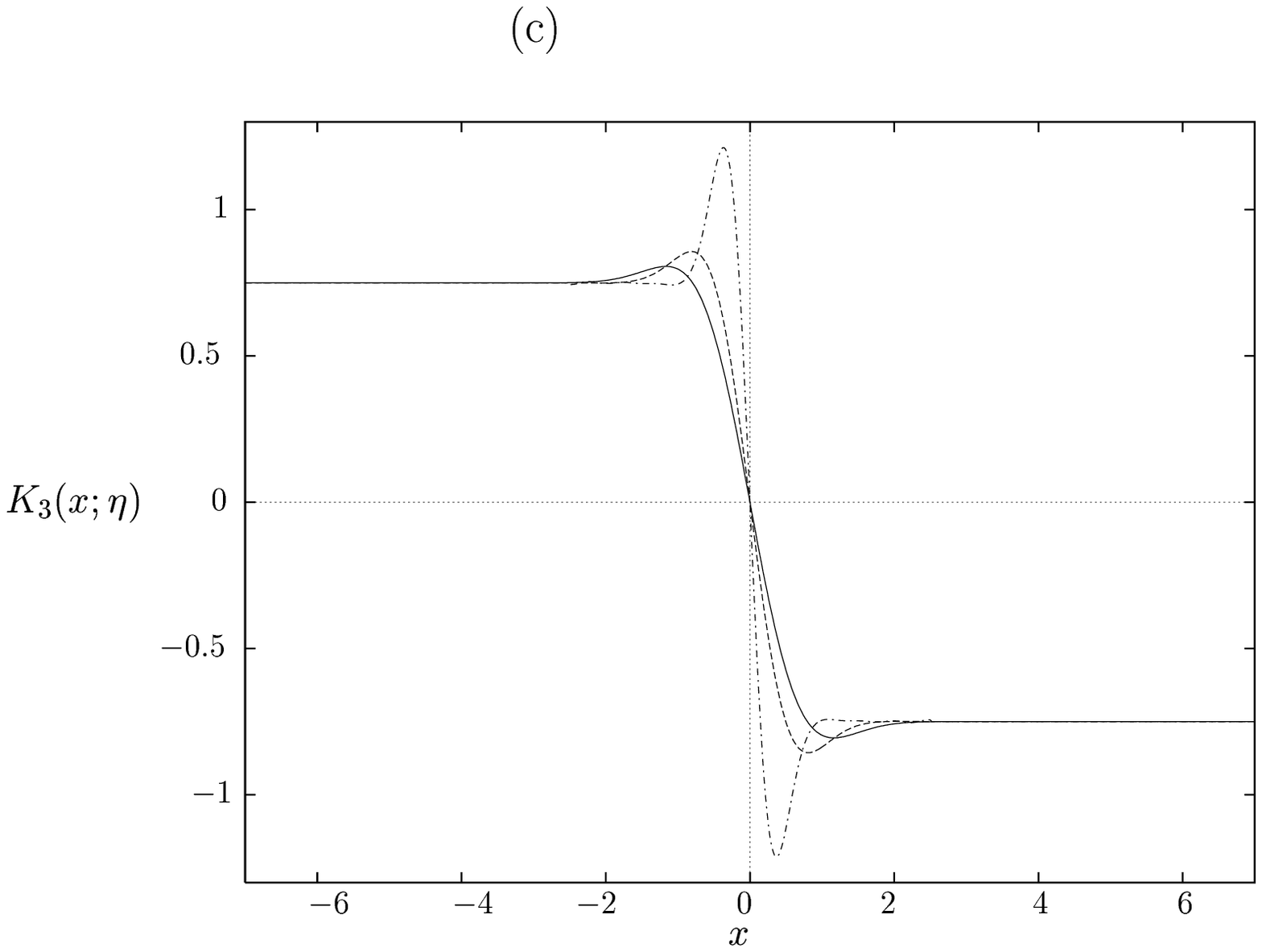,width=\linewidth}
\end{minipage}\hfill
\begin{minipage}[b]{0.45\linewidth}
\centering\epsfig{figure=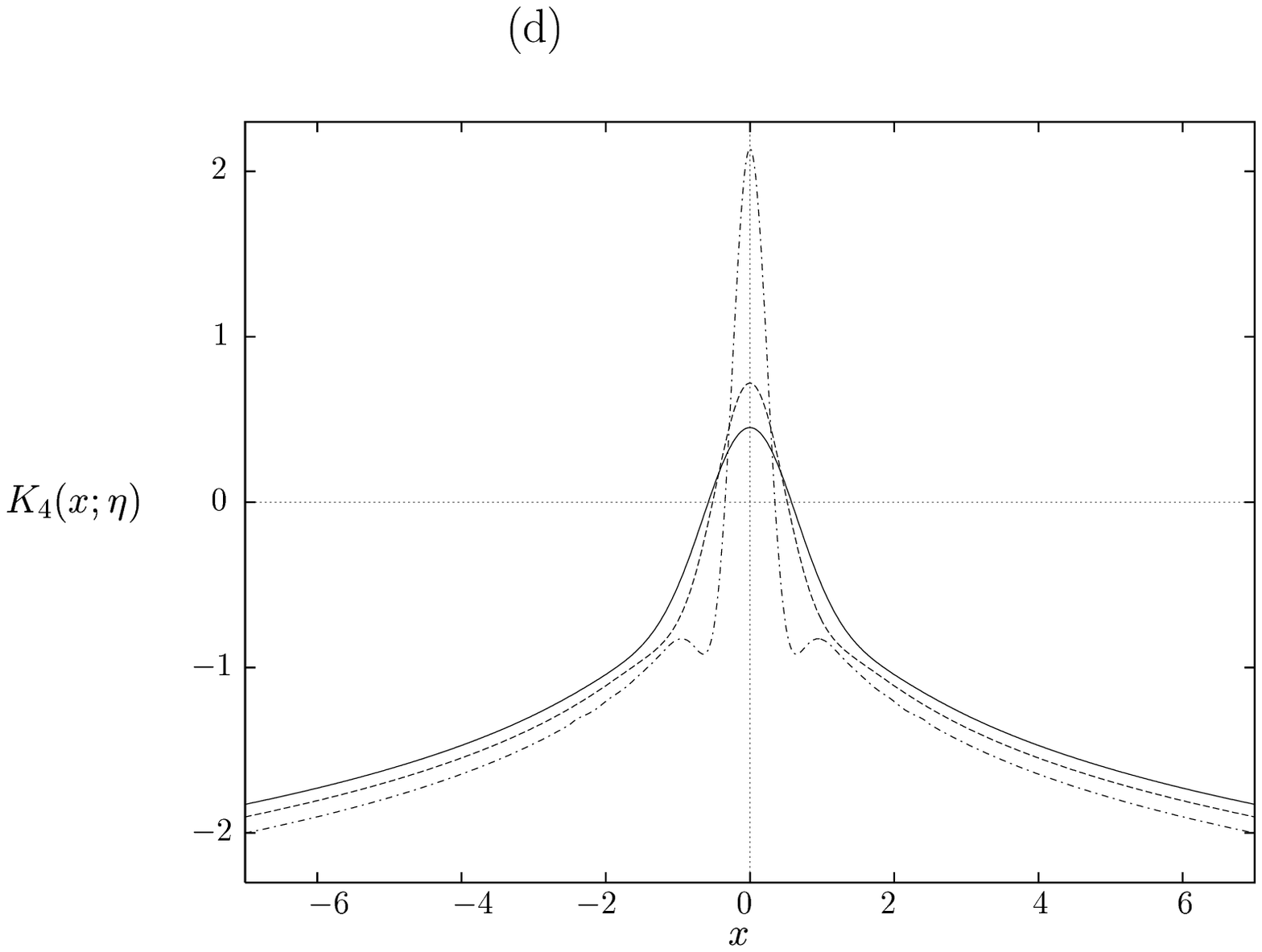,width=\linewidth}
\end{minipage}
\caption{
The $x$-dependent part $K_{k}(x;\eta)$
of the sampling function $K_{k}(x,\vartheta;\eta)$
$\!=$ $\!e^{ik\vartheta} K_{k}(x;\eta)$ for the determination 
of the exponential phase moments $\Psi_{k}$ from the smeared
quadrature-component distribution $p(x,\vartheta;\eta)$ 
is shown for various $k$ and $\eta$ [$\eta$ $\!=$ $\!1$ (full lines),
$\eta$ $\!=$ $\!0.8$ (dashed lines), $\eta$ $\!=$ $\!0.6$ 
(dashed-dotted lines)]. 
\label{F3}
}
\end{figure}

It is worth noting that under certain circumstances it is possible
to compensate for the systematic error during the sampling process,
introducing an appropriately modified kernel $K_{k}(x;\eta)$. 
Let us again assume Gaussian smearing, which is typically
observed in nonperfect detection, and apply Eq.~(\ref{smear1})
together with Eq.~(\ref{smear2}). In this case we may replace 
Eq.~(\ref{Ma1}) with \cite{DAriano}
\begin{eqnarray}
\lefteqn{
\langle \hat{a}^{\dagger n}\hat{a}^m\rangle
\!=\! \left[2\pi \sqrt{(2\eta)^{n+m}} {n\!+\!m\choose m}\right]^{-1}
}
\nonumber \\ && \hspace{10ex} \times
\int_{2\pi} d \vartheta  \int_{-\infty}^{\infty} dx \, 
e^{i(n-m)\vartheta} \, {\rm H}_{n+m}(x)
\,p(x,\vartheta;\eta),
\label{los1}
\end{eqnarray}
and follow the lines given in Sec.~\ref{S3} and \ref{app0}. 
It is easily seen that in Eq.~(\ref{Ma1xa}) 
$p(x,\vartheta)$ and $\tilde K_{k}(x)$, respectively, must be replaced 
with $p(x,\vartheta;\eta)$ and $\tilde K_{k}(x;\eta)$, provided that
$\tilde K_{k}(x;\eta)$ exists. The kernel $\tilde K_{k}(x;\eta)$ 
obviously compensates for the losses associated with nonperfect detection 
and can be obtained from Eq.~(\ref{05}) in \ref{app0}, if $C_{l}^{(k)}$ 
is replaced with $C_{l}^{(k)}(\eta)$ $\!=$ $\!\eta^{-(l+k/2)}C_{l}^{(k)}$.
In close analogy to Eq.~(\ref{kt1}) we then find that the modified 
kernel $\tilde K_{k}(x;\eta)$ can be rewritten as
\begin{eqnarray}
\label{los3}
\tilde K_{k}(x;\eta)= K_{k}(x;\eta) - {\rm F}_{k}(x;\eta),
\end{eqnarray}
where for even and odd $k$, respectively, $\eta^{k/2}K_{k}(x;\eta)$ is 
given by Eqs.~(\ref{kernev}) and (\ref{kernod}), if in the integrals
$z_{k}$ is replaced with $z_{k}(\eta)$ $\!=$ $\!z_{k}/\eta$. Finally,
Eqs.~(\ref{A1}) and (\ref{A2}), respectively, are replaced with 
\begin{eqnarray}
\lefteqn{
K_{2m+1}(x;\eta) = (-1)^m \, \frac{ 2x (m\!+\!1)! }{ (2\pi/\eta)^{m+3/2} }
\int_{0}^{+\infty} dr \, 
\Bigg\{
\Omega^{(2m+1)}(r^2)
}
\nonumber \\ && \hspace{15ex} \times \,
\frac{ r^{2m}  
\Phi[m\!+\!2,3/2,-x^2\lambda(r^{2};\eta)\tanh(r^2/2)]}
{e^{-(m+1)r^2} \sinh^m(r^2/2) [\lambda(r^{2};\eta)\cosh(r^2/2)]^{m+2}}
\Bigg\} 
\label{A1eta}
\end{eqnarray}  
and
\begin{eqnarray}
\lefteqn{
K_{2m}(x;\eta) = (-1)^m \, \frac{m!}{(2\pi/\eta)^{m+1}}
\int_{0}^{+\infty} dr \,
\Bigg\{
\Omega^{(2m)}(r^2)
}
\nonumber \\ && \hspace{5ex} \times \,
\frac{ r^{2m-1} e^{m r^2 / 2} }
{ \sinh^m(r^2/2) }
\bigg[
\frac{ \Phi[m\!+\!1, 1/2,- x^2 \lambda(r^{2};\eta)\tanh(r^2/2)] }
{ e^{-(m+1)r^2/2} [\lambda(r^{2};\eta)\cosh(r^2/2)]^{m+1} }-1
\bigg]
\Bigg\} ,
\label{A2eta}
\end{eqnarray}
where
\begin{eqnarray}
\label{etafactor}
\lambda(r^{2};\eta) = 1 + (\eta - 1)(1 + e^{-r^{2}})^{-1},
\end{eqnarray} 
and the polynomial ${\rm F}_k(x;\eta)$ reads as
\begin{eqnarray}
\lefteqn{
{\rm F}_k(x;\eta) = 
\frac{1}{2\pi (2\eta)^{k/2}}
\sum_{n=1}^{[\frac{k-1}{2}]}
\Bigg[
\frac{(-2\eta)^n(k-n)!}{(k-2n)!} \, {\rm H}_{k-2n}(x)
}
\nonumber \\ && \hspace{15ex} \times\,
\sum_{l=0}^{\infty}{n\!+\!l\!-1\choose l}
\frac{1}{\sqrt{(l+1)\dots(l+k)}} 
\Bigg] .
\label{los6}
\end{eqnarray}
Needless to say that the polynomial can again be omitted since both 
$\tilde K_{k}(x;\eta)$ and $K_{k}(x;\eta)$ are solutions of the problem
and in practical measurements $K_{k}(x;\eta)$ is more suited for error 
reduction than $\tilde K_{k}(x;\eta)$. The numerical 
calculation of $K_{k}(x;\eta)$ can be performed in a way as outlined
in Sec.~\ref{S3} for $K_{k}(x)$. Examples of $K_{k}(x;\eta)$ for various 
values of $k$ and $\eta$ are shown in Fig.~\ref{F3}.

It should be pointed out 
that the sum rules (\ref{Ma8}) and (\ref{Ma82}) used in the derivation
only apply when $|z_{k}(\eta)|$ $\!<$ $\!1$. Hence we observe that the 
condition $\eta$ $\!>$ $\!1/2$ must be fulfilled in order to compensate
for Gaussian data smearing, which is analogous to the 
density matrix reconstruction in the Fock basis \cite{Ariano1}. It is 
worth noting that the condition $\eta$ $\!>$ $\!1/2$ corresponds to the 
requirement that the width of the Gaussian (\ref{smear2}) is smaller 
than the vacuum noise. From Fig.~\ref{F3}
we see that with increasing $|x|$ the kernel
$K_{k}(x;\eta)$ for odd $k$ rapidly approaches the classical limit  
(\ref{cl4}) for perfect detection, whereas for even $k$ 
it approaches the classical limit 
(\ref{cl5}) up to an irrelevant $\eta$-dependent constant. The results 
reveal that in classical optics it is impossible to compensate for
the losses in nonperfect detection, because of the vanishing 
vacuum noise of a classical oscillator. 
  
As expected, substantial differences between $K_{k}(x;\eta)$ and
$K_{k}(x)$ are observed in the region around $x$ $\!=$ $\!0$,
and they increase with decreasing $\eta$ (Fig.~\ref{F3}). 
The [compared with $K_{k}(x)$] stronger variation of $K_{k}(x;\eta)$ 
implies that the use of $K_{k}(x;\eta)$ for sampling of the
exponential phase moments from the smeared quadrature-component
distribution gives rise to a larger statistical error than the use
of $K_{k}(x)$ (for the statistical error, see Sec. \ref{Ser}).
This is obviously the price paid for suppression of the systematic
error. Based on the precision of the data available, the experimenter should 
therefore decide whether to use $K_{k}(x;\eta)$ (which increases the 
statistical error) or $K_{k}(x)$ (which decreases the statistical error
but introduces a bias). 


\subsection{Phase discretization}
\label{phasediscret}

In practice, $p(x,\vartheta)$ can only be measured at $N$ discrete phases
$\vartheta_{l}$. When they are equidistantly distributed
over a $2\pi$ interval, i.e., $\vartheta_{l}$ $\!=$ $\! (2\pi/N) l$,
where $l$ $\!=$ $\!0,1,\dots N\!-\!1$, then application of Eq.~(\ref{c6})
yields the experimentally determined exponential phase moments
\begin{eqnarray}
\label{ed2}
\Psi_k(N)
= \frac{2\pi}{N} \sum_{l=0}^{N-1} e^{ik\vartheta_{l}}
\int_{-\infty}^{\infty} dx \, K_{k}(x) \, p(x,\vartheta_{l}) ,
\end{eqnarray}
which can be rewritten as, on using Eq.~(\ref{qi3}), 
\begin{eqnarray}
\label{ed3}
\Psi_k(N)
= \frac{2\pi}{N} \sum_{l=0}^{N-1} \sum_{m,n=0}^{\infty} 
e^{i\frac{2\pi}{N}(k\!+\!n\!-\!m)l} \varrho_{m,n}
 \int_{-\infty}^{\infty} \! dx K_{k}(x)
\psi_{m}(x) \psi_{n}(x) .
\end{eqnarray}
Taking into account that $N^{-1}\sum_{l=0}^{N-1} e^{i 2\pi (k+n-m)l /N}$
$\! =$ $\! \delta_{k+n-m \, {\rm mod}\, N}$ and recalling
Eqs.~(\ref{qi1}) and (\ref{qi5}), we derive 
\begin{eqnarray}
\label{ed5}
\Psi_k(N) = \Psi_k + \Delta\Psi_{k}^{\rm (d)} ,
\end{eqnarray}
where 
\begin{eqnarray}
\label{ed5a}
\Delta\Psi_k^{\rm (d)}
= \sum_{s=1}^{\infty} \sum_{n=0}^{\infty} 
\left(
\varrho_{n+k+sN,n} Q_{n+k+sN,n}^{(k)}
+ \varrho_{n,n+sN-k} Q_{n,n+sN-k}^{(k)}
\right)
\end{eqnarray}
represents the systematic error owing
to phase discretization. In Eq.~(\ref{ed5a}) the abbreviation
\begin{eqnarray}
\label{ed6}
Q_{m,n}^{(k)} = 2\pi \int_{-\infty}^{\infty} dx \, K_{k}(x)
\psi_{m}(x) \psi_{n}(x) 
\end{eqnarray}
is used and it is assumed, for notational convenience, that 
$N$ $\!>$ $\!k$. Note that from physical arguments it is also 
reasonable to assume that the number of phases is larger than the 
index of the measured moment (otherwise the systematic error 
could dominate the result).

   From Eq.~(\ref{ed5a}) we see that the error is influenced by all 
off-diagonal density-matrix elements of the type of $\varrho_{n+k\pm sN,n}$.
The effect, which is also called ``aliasing'', has also been
found in the reconstruction of the density matrix in the Fock basis 
from the data measured in balanced \cite{aliasing} and unbalanced 
\cite{unbalancedJMO} homodyning.
For highly excited states (i.e., $\hat \varrho_{n,m}$ $\!\approx$ 
$\!0$ if $n,m$ $\!<$ $\!n_{0}$, with \mbox{$n_{0}$ $\!\gg$ $\!1$)} the 
relevant $Q_{m,n}^{(k)}$ can be approximately calculated, using in 
Eq.~(\ref{ed6}) the classical kernel given in Eqs.~(\ref{cl4}) 
and (\ref{cl5}):
\begin{eqnarray}
\label{ed7}
Q_{n+k+sN,n}^{(k)} \approx (-1)^{Ns/2} \frac{k}{sN\!+\!k} \,, 
\quad
Q_{n,n+sN-k}^{(k)} \approx - Q_{n-k+sN,n}^{(k)},
\end{eqnarray}
Note that $sN$ is even, since from the symmetry properties of 
$Q_{m,n}^{(k)}$ it follows that $Q_{m,n}^{(k)}$ $\!=$ $\!0$ if 
$m$ $\!+$ $\!n$ $\!+$ $\!k$ is odd. Combining Eqs.~(\ref{ed5a})
and (\ref{ed7}) yields 
\begin{eqnarray}
\label{ed9}
\Delta\Psi_k^{\rm (d)} \approx 
\sum_{s=1}^{\infty}
(-1)^{s}
\left( \frac{k}{2sN\!+\!k}
\Psi_{k\!+\!2Ns}
+ \frac{k}{2sN\!-\!k}
\Psi_{k\!-\!2Ns}
\right)
\end{eqnarray}
for odd $N$ and
\begin{eqnarray}
\label{ed8}
\Delta\Psi_k^{\rm (d)} \approx 
\sum_{s=1}^{\infty}
(-1)^{Ns/2}
\left( \frac{k}{sN\!+\!k}
\Psi_{k\!+\!Ns}
+ \frac{k}{sN\!-\!k}
\Psi_{k\!-\!Ns}
\right)
\end{eqnarray}
for even $N$. We see that the error of the measured 
exponential phase moment is expressed in terms of higher-order moments,
and it decreases with increasing $N$. Note that the difference between
the errors in Eqs.~(\ref{ed9}) and (\ref{ed8}) reflects the fact that with
regard to a $\pi$ interval, the number of different phases is twice 
as large for odd $N$ as for even $N$, which of course substantially reduces
the systematic error in the first case.   


\subsection{Statistical error}
\label{Ser}

When in an experiment $n(\vartheta_{l})$ measurements are performed for 
each phase $\vartheta_{l}$, the exponential phase moments can be estimated
as, on applying Eq.~(\ref{c6}),
\begin{eqnarray}
\label{es1}
\Psi_k^{\rm (est)}(N)
= \frac{2\pi}{N} \sum_{l=0}^{N-1} e^{ik\vartheta_{l}}
\,\frac{1}{n(\vartheta_{l})}
\sum_{r=1}^{n(\vartheta_{l})}
K_{k}[x_{r}(\vartheta_{l})] ,
\end{eqnarray}
where $x_{r}(\vartheta_{l})$ is the result of the $r$th 
individual measurement at the phase $\vartheta_{l}$.
Taking the average of all estimates $\Psi_k^{\rm (est)}(N)$, 
\begin{eqnarray}
\label{es2}
\overline{ \Psi_k^{\rm (est)}(N) } = 
\frac{2\pi}{N} \sum_{l=0}^{N-1} e^{ik\vartheta_{l}}
\, \frac{1}{n(\vartheta_{l})}
\sum_{r=1}^{n(\vartheta_{l})} \overline{
K_{k}[x_{r}(\vartheta_{l})] } ,
\end{eqnarray}
yields, as expected, $\Psi_{k}(N)$ from Eq.~(\ref{ed2}),
\begin{eqnarray}
\label{es4}
\overline{ \Psi_k^{\rm (est)}(N) } = {\Psi_k^{(N)}} ,
\end{eqnarray}
because of
\begin{eqnarray}
\label{es3}
\overline{ K_{k}(x_{r}(\vartheta_{l})) } = 
\int_{\infty}^{\infty} dx \, K_{k}(x) \, p(x,\vartheta_{l}) .
\end{eqnarray}
The variances of the real and imaginary parts of $\Psi_k^{\rm (est)}(N)$
can be obtained in a similar way. Taking into account that the
individual measurements are independent of each other, we derive
\begin{eqnarray}
\label{es5}
{\rm Var}\!\left\{ {\rm Re}[ \Psi_k^{\rm (est)}(N) ] \right\} 
= \frac{4\pi^{2}}{N^{2}} \sum_{l=0}^{N-1}
\frac{\cos ^{2}(k\vartheta_{l})}{n(\vartheta_{l})}
\,\overline{ \left\{\Delta K_{k}[x_{r}(\vartheta_{l})]\right\}^{2} } ,
\end{eqnarray}
and
\begin{eqnarray}
\label{es6}
{\rm Var}\!\left\{ {\rm Im}[ \Psi_k^{\rm (est)}(N) ] \right\} 
= \frac{4\pi^{2}}{N^{2}} \sum_{l=0}^{N-1}
\frac{\sin ^{2}(k\vartheta_{l})}{n(\vartheta_{l})}
\,\overline{ \left\{\Delta K_{k}[x_{r}(\vartheta_{l})]\right\}^{2} } ,
\end{eqnarray}
where
\begin{eqnarray}
\label{es7}
\overline{ \left\{\Delta K_{k}[x_{r}(\vartheta_{l})]\right\}^{2} } 
= \int_{-\infty}^{\infty} dx \, K_{k}^{2}(x) \, p(x,\vartheta_{l}) 
- \left[
\int_{-\infty}^{\infty} dx \, K_{k}(x) \, p(x,\vartheta_{l})
\right]^{2} .
\end{eqnarray}

Equations (\ref{es5}) -- (\ref{es7})
enable us to estimate the statistical error of the measured moments,
substituting in Eqs.~(\ref{es5}) and (\ref{es6}) for 
$\overline{ \left\{\Delta K_{k}[x_{r}(\vartheta_{l})]\right\}^{2} }$
the corresponding estimates. From Eqs.~(\ref{es5}) -- (\ref{es7})
it is seen that the statistical error depends on the 
shape of the function $K_{k}(x)$. In order to reduce the statistical 
error, the ambiguity in the determination of $K_{k}(x)$ can be 
advantageously used to choose it such that it varies as slowly as possible.    
This is one of the reasons for omitting the polynomial in Eq.~(\ref{kt1}).
Moreover, the statistical error can be reduced when the number
of events, $n(\vartheta_{l})$, is appropriately varied with the phase
$\vartheta_{l}$. From Eqs.~(\ref{es5}) and (\ref{es6}) it is suggested 
to increase $n(\vartheta_{l})$ for such phases for which
$\overline{ \left\{\Delta K_{k}[x_{r}(\vartheta_{l})]\right\}^{2} }$,
Eq.~(\ref{es7}), becomes relatively large. This is typically the case
when $p(x,\vartheta)$ is essentially nonzero in an $x$ interval 
around $x$ $\!=$ $\!0$, in which $K_{k}(x)$ strongly varies with $x$. 
Note that this result is in qualitative agreement with that of the
maximum-likelihood method for estimating phase shifts \cite{Hradil}.


\section{Computer simulations of measurements}
\label{S5}
\subsection{Exponential phase moments}
\label{S5a}

\begin{figure}
\centering\epsfig{figure=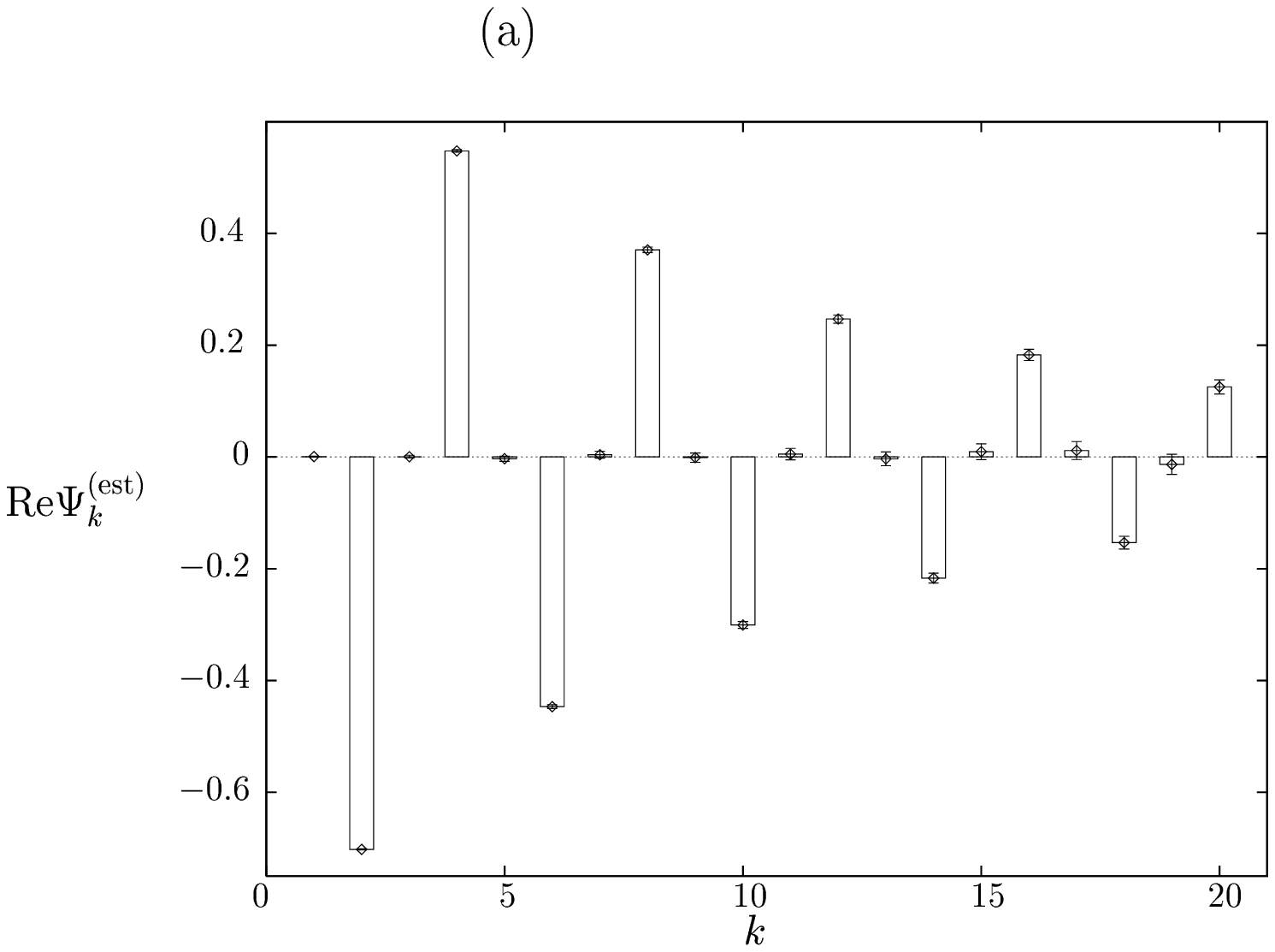,width=0.7\linewidth}

~
\vspace{0.1cm}

~

\centering\epsfig{figure=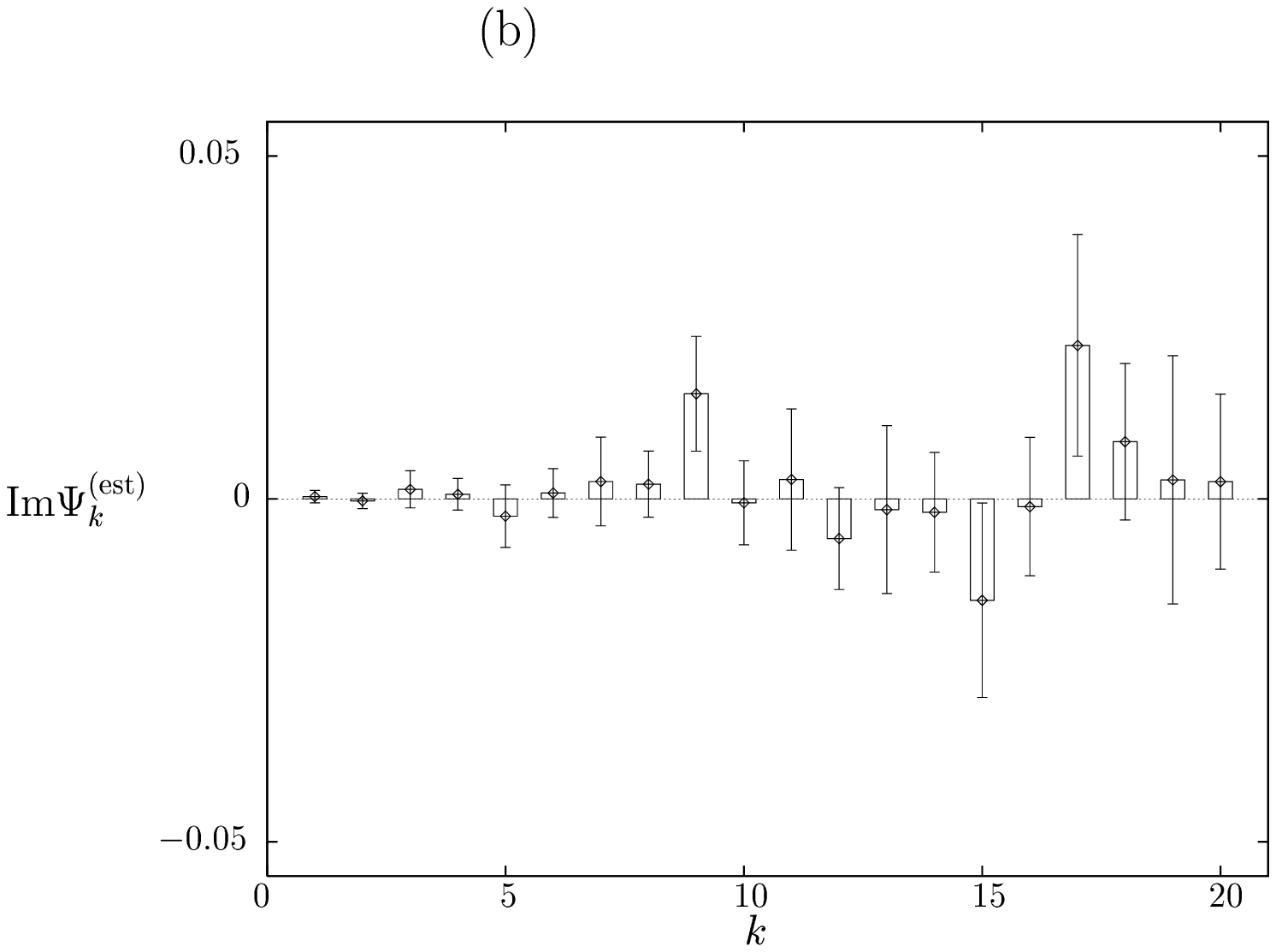,width=0.7\linewidth}
\caption{
Examples of measured exponential phase moments $\Psi_{k}^{\rm (est)}$ 
are shown for a mode prepared
in a squeezed vacuum $|0\rangle_s$ $\!=$ $\!\hat{S}(\xi)|0\rangle$,
with $\xi$ $\!=$ $\!-1.31$, i.e., $\langle\hat{n}\rangle$ $\!=$ $\!3$
[(a) real part of $\Psi_{k}^{\rm (est)}$; (b) imaginary part of 
$\Psi_{k}^{\rm (est)}$]. The
error bars indicate the estimated statistical error. 
In the computer simulation it is assumed that the quadrature component 
distribution $p(x,\vartheta)$ is detected at $120$ phases $\vartheta$ 
equidistantly distributed in a $2\pi$ interval and that at each phase
$10^{4}$ events are recorded.
\label{F4}
}
\end{figure}

\begin{figure}
\centering\epsfig{figure=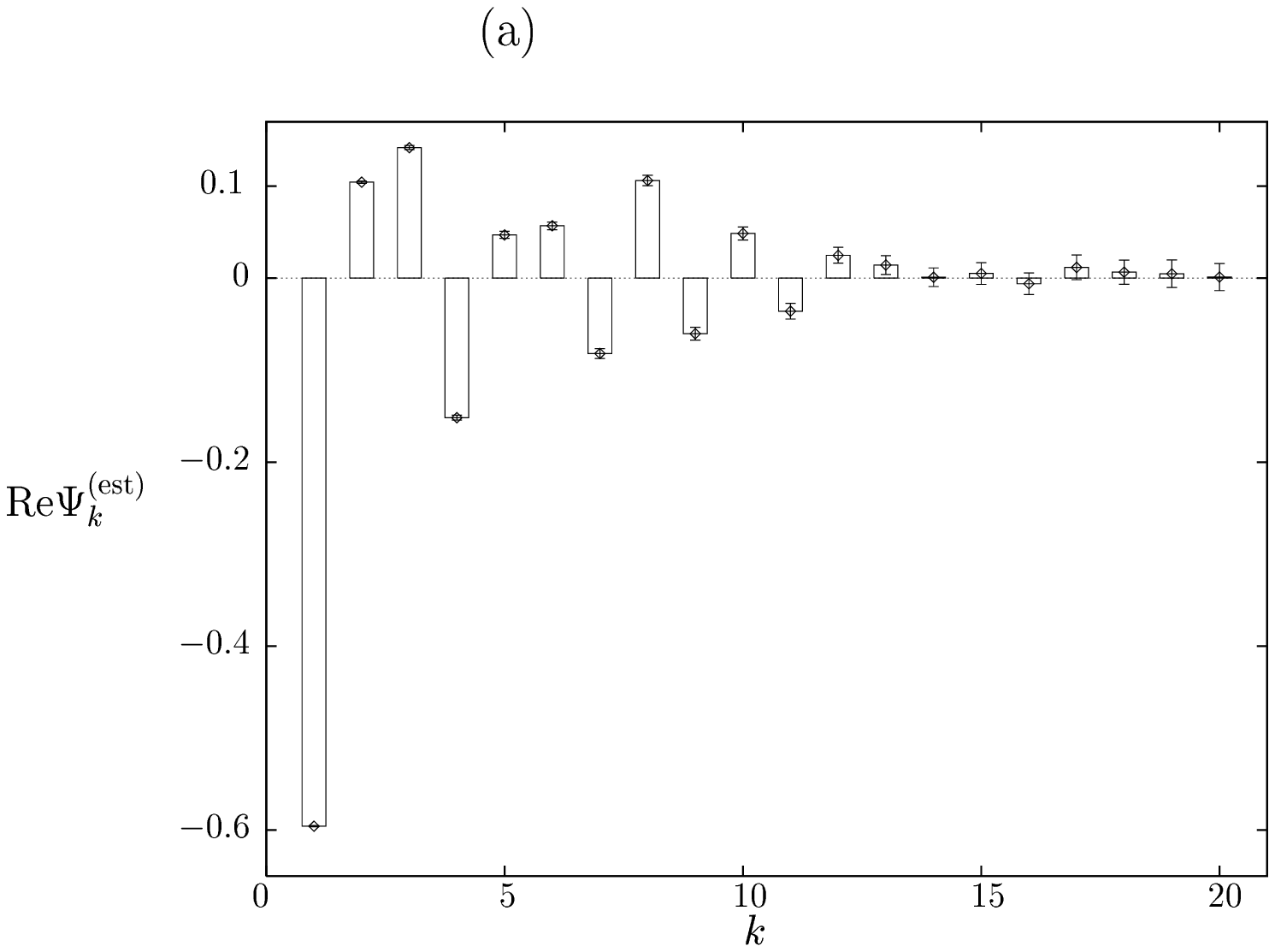,width=0.7\linewidth}

~
\vspace{0.1cm}

~

\centering\epsfig{figure=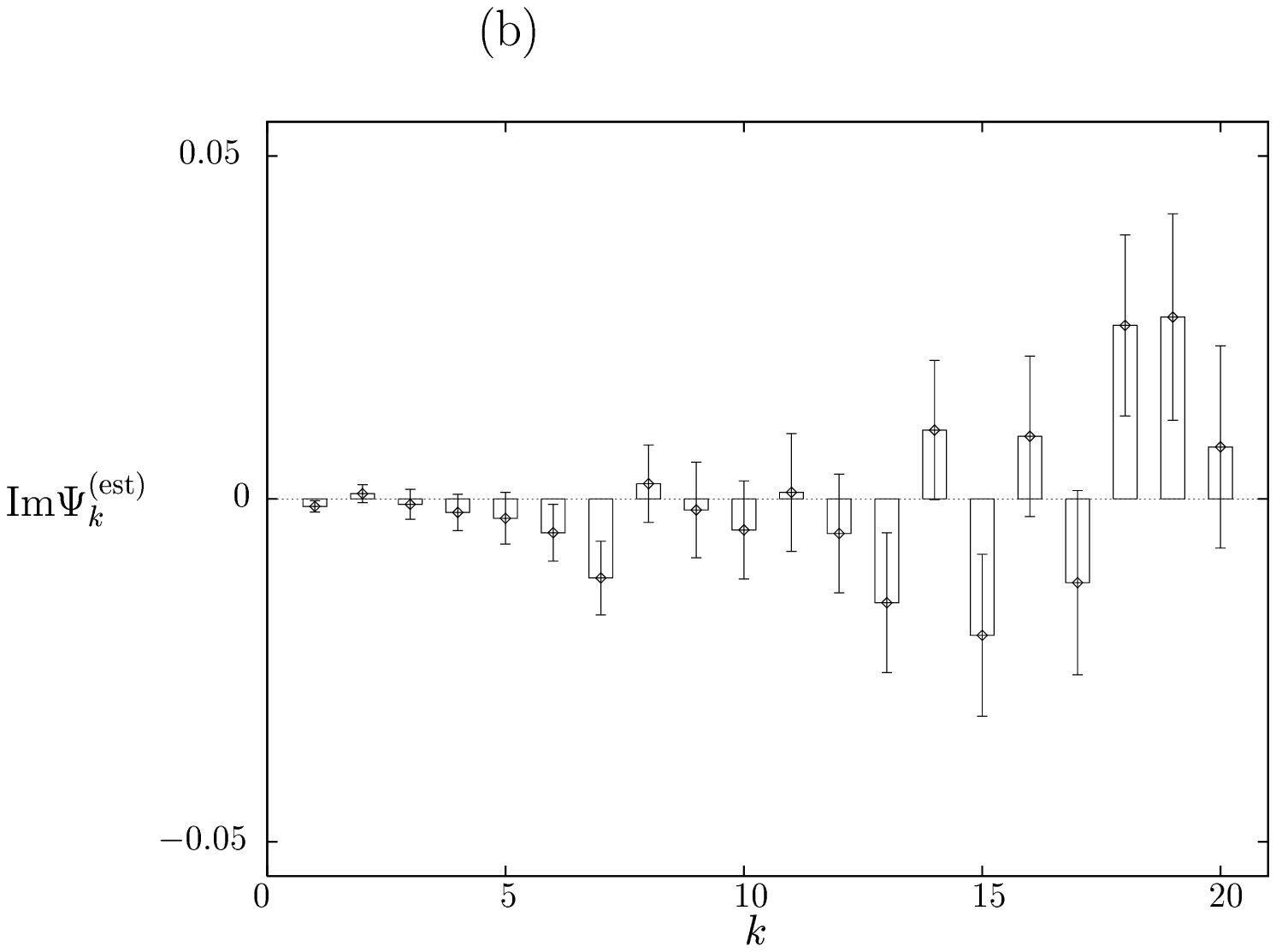,width=0.7\linewidth}
\caption{
Examples of measured exponential phase moments $\Psi_{k}^{\rm (est)}$ 
are shown for a mode prepared
in a displaced Fock state $|\alpha,n\rangle$ $\!=$
$\!\hat{D}(\alpha)|n\rangle$,
with $\alpha$ $\!=$ $\!-1.5$ and
$n$ $\!=$ $\!2$, i.e., $\langle\hat{n}\rangle$ $\!=$ $\!4.25$
[(a) real part of $\Psi_{k}^{\rm (est)}$; (b) imaginary part of 
$\Psi_{k}^{\rm (est)}$]. The
error bars indicate the estimated statistical error. 
In the computer simulation it is assumed that the quadrature component 
distribution $p(x,\vartheta)$ is detected at $120$ phases $\vartheta$ 
equidistantly distributed in a $2\pi$ interval and that at each phase
$10^{4}$ events are recorded.
\label{F5}
}
\end{figure}

To illustrate the method, we have performed computer simulations of
measurements of the quadrature component distribution $p(x,\vartheta)$,
assuming the signal field to be prepared in various states, such as 
a squeezed vacuum $|0\rangle_{\rm s}$ $\!=$ $\!\hat{S}(\xi)|0\rangle$ 
$\!=$ $\!\exp\{- \frac{1}{2} [ \xi (\hat a ^\dagger)^2 - 
\xi^\ast \hat a^2 ] \} |0\rangle $ and a displaced Fock state
$|\alpha,n\rangle$ $\!=$ $\!\hat{D}(\alpha)|n\rangle$ $\!=$
$\!\exp( \alpha\hat a^\dagger-\alpha^\ast\hat a ) |n\rangle $.
We have restricted attention to perfect detection and assumed
that the measurements are performed at $N$ $\!=$ $\!120$ (equidistant)
phases $\vartheta_{l}$ within a $2\pi$ interval and $n(\vartheta_{l})$
$\!=$ $10^4$ events are recorded at each phase.
Examples of the sampled exponential phase moments $\Psi_{k}$ are 
shown in Figs.~\ref{F4} and \ref{F5}
for a squeezed vacuum and a displaced Fock state, respectively. 
The error bars indicate the standard deviations obtained according to 
Eqs.~(\ref{es5}) and (\ref{es6}). Compared to the statistical error,
the systematic error due to phase discretization is (for the
chosen number $N$ of phases $\vartheta_{l}$) negligible small.
   From Figs.~\ref{F4} and \ref{F5} we see that [for the chosen numbers
$n(\vartheta_{l})$ of events] the exponential phase moments are 
obtained with sufficiently good accuracy, provided that $k$ is small
enough. We further see that the accuracy decreases with increasing $k$. 
(Note that for the chosen state parameters the imaginary parts 
must vanish for all $k$.)
Clearly, the accuracy can be improved if the number of measurements
is increased.    


\subsection{Phase distribution}
\label{S5b}

The possibility of direct sampling of exponential phase
moments $\Psi_{k}$ offers novel possibilities of experimental 
verification of fundamental number--phase uncertainty relations, as has
been shown recently \cite{Opatrny1}. It is worth noting
that the measurements can be performed with high precision
since only low-order moments play a role.
Here we address the problem of the determination of the
whole phase distribution $P(\varphi)$. 

\begin{figure}
\centering\epsfig{figure=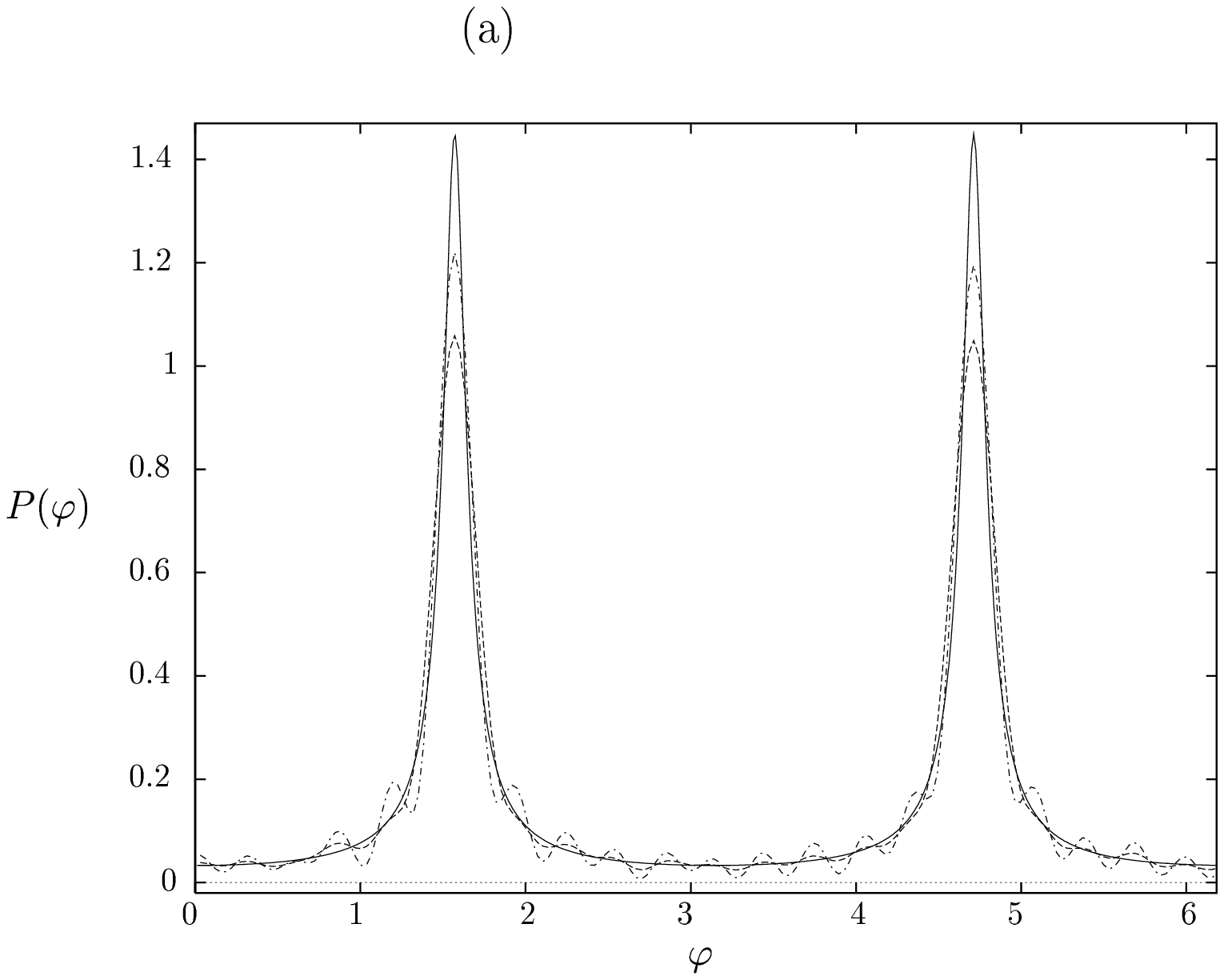,width=0.7\linewidth}

~
\vspace{0.1cm}

~

\centering\epsfig{figure=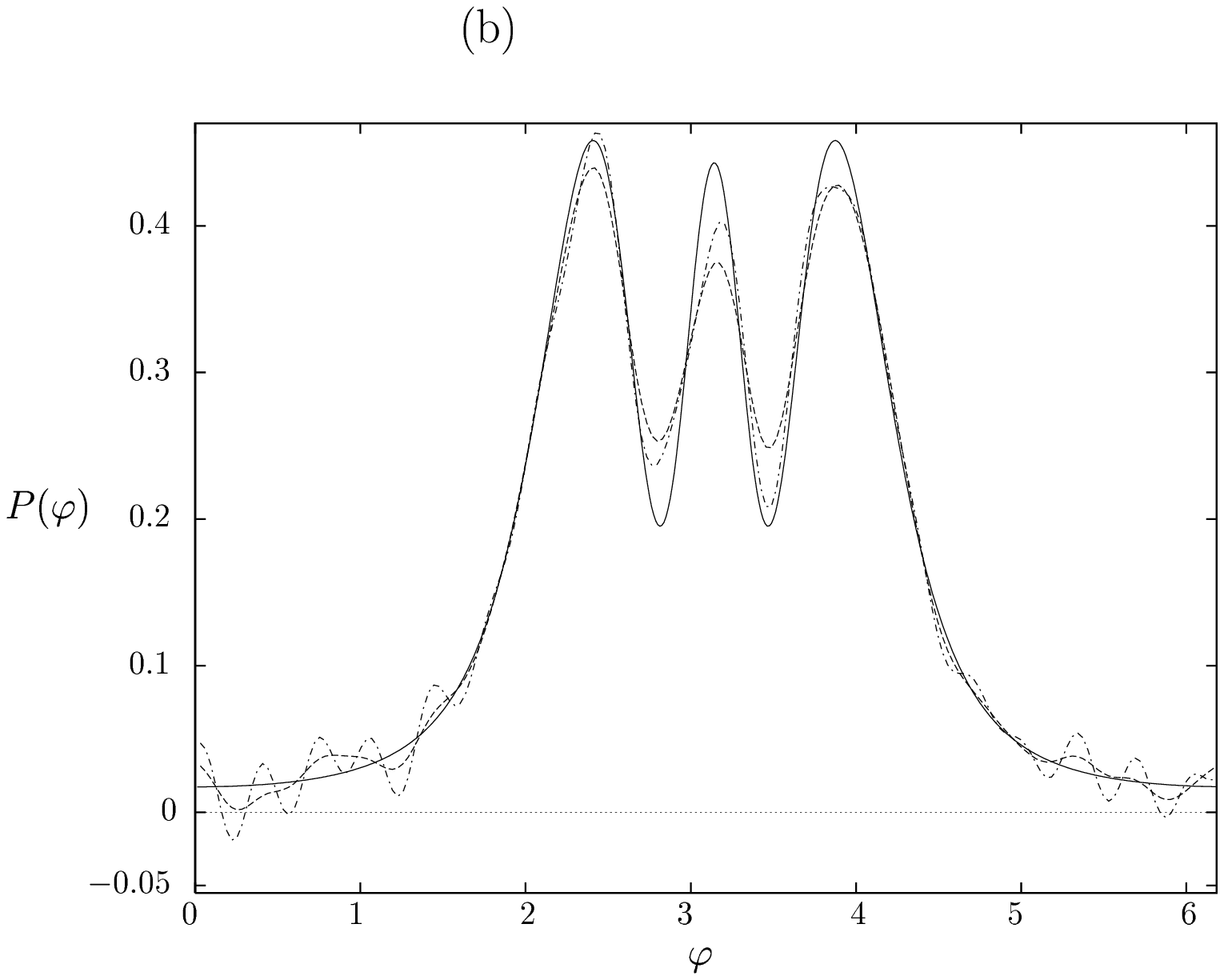,width=0.7\linewidth}
\caption{
The canonical phase distribution $P(\varphi)$ reconstructed from
$K$ $\!=$ $\!20$ measured exponential phase moments 
$\Psi_{k}^{\rm (est)}$ given
in Figs.~\protect\ref{F4} and \protect\ref{F5} is shown
for the squeezed vacuum (a) and the displaced Fock state (b)
therein. The results of direct application of Eq.~(\protect\ref{recophase1})
(dashed-dotted lines) and application of regularized least-squares inversion
(dashed lines) are compared with the exact distributions (solid lines). 
\label{F6}
}
\end{figure}

Since the exponential phase 
moments are nothing but the Fourier components of the phase distribution
[Eq.~(\ref{c4})], the sampled moments can be used to reconstruct
the phase distribution according to
\begin{equation}
P(\varphi)=\frac{1}{2\pi}\sum_{k=-\infty}^{\infty} e^{-ik\varphi} \, \Psi_k .
\label{recophase1}
\end{equation} 
Moreover, since any physical quantum state can be approximated to any desired
degree of accuracy by truncating it at some photon number $n_{\rm max}$
if $n_{\rm max}$ is suitably large, from Eq.~(\ref{qi1}) it follows
that (for chosen accuracy) the number of moments $\Psi_{k}$ that effectively
contribute to $P(\varphi)$ in Eq.~(\ref{recophase1}) is finite, 
i.e., $|k|$ $\!=$ $\!1,2,\ldots,K$, with $K$ $\!=$ $\!n_{\rm max}$.      
Hence, $P(\varphi)$ can be obtained truncating the sum
in Eq.(\ref{recophase1}) at $|k|$ $\!=$ $\!K$ and substituting 
for the $\Psi_{k}$ the measured moments $\Psi_{k}^{\rm (est)}$. The phase 
distributions that are reconstructed from the measured moments given in 
Figs.~\ref{F4} and \ref{F5} for a squeezed vacuum and a displaced Fock 
state, respectively, are plotted in Fig.~\ref{F6}, 
on assuming that $n_{\rm max}$ $\!=$ $\!20$. The statistical error
of $\Psi_{k}^{\rm (est)}$ gives of course rise to an error of $P(\varphi)$.
Since the error in  $P(\varphi)$ can be obtained easily from the law of error 
propagation \cite{Bevington} in a standard way, we renounce the calculation here. 
  
Finally, it should be pointed out that there are other methods, 
such as least-squares inversion \cite{numrecip,Bevington} and 
maximum-entropy inversion \cite{Jaynes1}, which can be used 
for reconstructing the phase distribution from the
measured (i.e., inaccurate) exponential phase moments -- methods
that have been successfully applied in various fields of physics.
Let us briefly comment on the application of the method of
least-squares inversion. For this purpose we
return to Eq.~(\ref{c4}) and ask for $P(\varphi)$ that 
best fits the experimental data at $M$ chosen phases  
$\!\varphi_{m}$ ($m$ $\!=$ $\!0,1,\ldots,M\!-\!1$, with $M$ $\!\gg$ $\!2K$).
An answer can be given applying the method of least-squares
inversion \cite{numrecip,Bevington} to the set of $2K$
linear equations for $M$ unknown $P(\varphi_{m})$ 
($\varphi_{m}$ $\!=$ $\!2\pi m/M$),
\begin{eqnarray}
&& {\rm Re}\,\Psi_{k} = \frac{2\pi}{M} 
\sum_{m=0}^{M-1} \cos (k\varphi_{m}) \, P(\varphi_{m}) ,
\label{recophase2} \\ 
&& {\rm Im}\,\Psi_{k} = \frac{2\pi}{M} 
\sum_{m=0}^{M-1} \sin (k\varphi_{m}) \, P(\varphi_{m}) 
\label{recophase3}
\end{eqnarray} 
($k$ $\!=$ $\!1,2,\ldots,K$), i.e., minimizing the functional
\begin{eqnarray}
\chi^2=
\lefteqn{
\sum_{k=1}^{K}
\left\{
\left[\sigma_k^{(\rm Re)}\right]^{-2}
\left[
{\rm Re}\,\Psi_k^{\rm (est)} 
- \frac{2\pi}{M} \sum_{m=0}^{M-1} P(\varphi_m) \cos(k\varphi_m) 
\right]^2
\right.
}
\nonumber \\ && \hspace{6ex}
+\left.
\left[\sigma_k^{(\rm Im)}\right]^{-2}
\left[
{\rm Im}\,\Psi_k^{\rm (est)} 
- \frac{2\pi}{M} \sum_{m=0}^{M-1} P(\varphi_m) \sin(k\varphi_m),
\right]^2
\right\},
\label{recophase4}
\end{eqnarray}
where $\sigma_k^{(\rm Re)}$ and $\sigma_k^{(\rm Im)}$, respectively, 
represent the errors involved in the determination of 
${\rm Re}\,\Psi_k^{\rm (est)}$ and ${\rm Im}\,\Psi_k^{\rm (est)}$. 
In particular, when $\sigma_k^{(\rm Re)}$ $\!\approx$ $\!\sigma_k^{(\rm Im)}$
$\!=$ $\!\sigma$ then (for \mbox{$M$ $\!\gg$ $\!2K$}) the resulting
$P(\varphi_{m})$ is in agreement with that obtained from
Eq.~(\ref{recophase1}), with $|k|$ $\!\leq$ $\!K$. From comparison
with the exact phase distribution it can be seen (Fig.~\ref{F6}) that
outside the regions in which the phase distribution is essentially nonzero
artificial oscillations and even negative values of the 
reconstructed distribution may be found, because of the statistical error
of the measured moments (cf. Figs.~\ref{F4} and \ref{F5}). The problem
can be partially overcome introducing regularization techniques
in the solution of Eqs.~(\ref{recophase2}) and (\ref{recophase3}) 
(for details, see \cite {numrecip}), as it can be seen from 
Fig.~\ref{F6}. The figure also reveals that the
artifacts are suppressed at the expense of a smearing of the
overall distribution.


\section{Summary and conclusions}
\label{S6}
We have presented a method for direct sampling of the exponential moments 
of the canonical phase of a single-mode radiation field from the data recorded 
in balanced homodyning. 
The sampling method enables us to determine the moments in real time,
together with the statistical error. It is worth noting that the method
renders it possible to closely relate the basic-theoretical concept of 
canonical phase to the experiment. 
The sampling functions relating the quadrature-component distribution to
the exponential phase moments are well behaved. With increasing 
quadrature-component they rapidly approach their classical counterparts 
given either by step functions (for odd moments) or logarithmic functions 
(for even moments). In this way the concept provides us with a unified 
approach to the experimental determination of the canonical phase in both 
quantum and classical optics.

In our approach to the construction of the kernel functions needed 
for direct sampling the exponential phase moments we have 
extended the proposal made in \cite{DAriano}. Hence the here used
subtraction of the polynomial arbitrariness from the sampling  functions can be
considered as a significant improvement. In fact the omittion of the ambiguity 
from the kernels
provided us with functions that can be universally used for any physical state
and that are much more  insensitive to errors.
This point should be also considered if a similar
approach to the derivation of the kernel corresponding to any other generic field 
operator is adopted. 

In order to study the accuracy of the method, we have discussed the 
influence of various experimental inaccuracies on the measured
exponential phase moments. In particular, the finite number of 
local-oscillator phases results in an aliasing effect, whereas smearing 
of the quadrature-component causes a bias toward smaller absolute
values of the moments. When the data smearing results from
imperfect detection with efficiency larger 50\% $\eta$, then 
modified sampling functions can be introduced for compensating the losses.

To illustrate the applicability of the method, we have performed
computer simulations of measurements for two quantum states and
and presented the measured exponential phase moments including
the statistical error. Finally, we have used the moments for a
reconstruction of the whole phase distribution. The results obtained
are in good agreement with the theoretical predictions.

\section*{Acknowledgments}
We are grateful 
to V. Pe\v{r}inov\'{a} for helpful and enlightening 
discussions. 
This work was supported by the Deutsche Forschungsgemeinschaft. 


\begin{appendix}
\renewcommand{\thesection}{Appendix \Alph{section}}

\section{Derivation of Eqs.~(\protect\ref{kt1}) -- (\protect\ref{Ma10})}
\label{app0}
\setcounter{equation}{0}
\renewcommand{\theequation}{\Alph{section} \arabic{equation}}

Combining Eqs.~(\ref{qic3}), (\ref{Ma3}), and (\ref{Ma1}) yields,
provided that the moments and correlations 
$\langle \hat{a}^{\dagger n}\hat{a}^m\rangle$ exist for all $n$ and $m$,
\begin{eqnarray}
\label{Ma0}
\Psi_{k} = \int_{2\pi} d\vartheta \int_{-\infty}^{\infty} dx 
\, \tilde K_{k}(x) e^{ik\vartheta} p(x,\vartheta) ,
\end{eqnarray}
where
\begin{equation}
\tilde K_{k}(x) =  (2\pi)^{-1} \sum_{l=0}^{\infty}
C_l^{(k)} \, {\rm H}_{2l+k}(x),
\label{05}
\end{equation}
with
\begin{eqnarray}
C_l^{(k)} = \frac{(l+k)!}{2^{l+k/2}(2l+k)}
\sum_{n=0}^l{\l\choose n} \, \frac{(-1)^{l-n}}
{\sqrt{(n+1)\dots (n+k)}} \, . 
\label{Ma5}
\end{eqnarray}
Using the relation
\begin{eqnarray}
\frac{1}{\sqrt{(n+1)\dots (n+k)}}
=\frac{1}{\pi^{k/2}}\int_{-\infty}^{+\infty}
dt_1 \, e^{-t_1^2} \, \cdots \,
\int_{-\infty}^{+\infty}
dt_k \, e^{-k{t_k}^2} e^{-nr_k^2} ,
\label{Ma6}
\end{eqnarray}
where 
\begin{eqnarray}
r_k^2 = \sum_{j=1}^{k} t_j^2 \, ,
\label{Ma6a}
\end{eqnarray}
we may rewrite Eq.~(\ref{Ma5}) as
\begin{eqnarray}
\lefteqn{
C_l^{(2m+1)}=
\frac{(2m+1+l)!}
{(2\pi)^{m+\frac{1}{2}} (2m+1+2l)!}
\int_{-\infty}^{+\infty} dt_1 \, e^{-t_1^2} \, \cdots \,
}
\nonumber \\ && \hspace{15ex} \times \, \cdots \,
\int_{-\infty}^{+\infty} dt_{2m+1} \, e^{-(2m+1)t_{2m+1}^2}
z_{2m+1}^l 
\label{Ma72}
\end{eqnarray}
and
\begin{eqnarray}
C_l^{(2m)}=\frac{(2m+l)!}
{(2\pi)^m (2m+2l)!}
\int_{-\infty}^{+\infty} dt_1 \, e^{-t_1^2} \, \cdots \,
\int_{-\infty}^{+\infty} dt_{2m} \, e^{-2mt_{2m}^2}
z_{2m}^l 
\label{Ma7}
\end{eqnarray}
for $k$ $\!=$ $\!2m\!+\!1$ and $k$ $\!=$ $\!2m$, respectively,
where
\begin{eqnarray} 
z_{k}  = {\textstyle\frac{1}{2}} (e^{-r_k^{2}}-1) . 
\label{Ma72aa}
\end{eqnarray}
We now substitute in Eq.~(\ref{05})
for $C_{l}^{(k)}$ the expressions \mbox{(\ref{Ma72})} and \mbox{(\ref{Ma7})}
and change the summation index $l$ as \mbox{$m$ $\!+$ $\!l$ $\!=$ $\!j$}.
In order to separate from $\tilde K_{k}(x)$ an irrelevant polynomial 
$F_{k}(x)$, we decompose $\tilde K_{k}(x)$ into two parts, 
\begin{eqnarray}
\tilde K_{k}(x) = K_{k}(x) - {\rm F}_{k}(x),
\label{kernel}
\end{eqnarray}
on rewriting the $j$ sum such that in F$_{k}(x)$ it runs from
$j$ $\!=$ $\!0$ and $j$ $\!=$ $\!1$, respectively, to \mbox{$j$ $\!=$ 
$\!m\!-\!1$} for odd and even $k$. In this way we find that 
\begin{eqnarray}
\lefteqn{
{K}_{2m+1}(x) = \frac{(m+1)!}{(2\pi)^{m+3/2}}
\int_{-\infty}^{+\infty} dt_1 \, e^{-t_1^2} \, \cdots \,
}
\nonumber \\ && \hspace{1ex} \times
\, \cdots \int_{-\infty}^{+\infty} dt_{2m+1} \, 
\frac{e^{-(2m+1)t_{2m+1}^2}}{z_{2m+1}^m}
\!\sum_{j=0}^{\infty}\frac{\Gamma(m\!+\!2\!+\!j)z_{2m+1}^j}
{(2j\!+\!1)!\Gamma(m\!+\!2)}
{\rm H}_{2j+1}(x) 
\label{Ma72a}
\end{eqnarray}
and
\begin{eqnarray}
\lefteqn{
{K}_{2m}(x) = \frac{m!}{(2\pi)^{m+1}}
\int_{-\infty}^{+\infty} dt_1 \, e^{-t_1^2} \, \cdots \,
}
\nonumber \\ && \hspace{1ex} \times 
\, \cdots \int_{-\infty}^{+\infty} dt_{2m} \, 
\frac{e^{-2mt_{2m}^2}}{z_{2m}^m}\!
\left[\sum_{j=0}^{\infty}
\frac{\Gamma(m\!+\!1\!+\!j)z_{2m}^j}
{(2j)!\Gamma(m\!+\!1)}{\rm H}_{2j}(x)
- 1\right] .
\label{Ma7a}
\end{eqnarray}
The polynomial $F_{k}(x)$ can be written as
\begin{eqnarray}
\lefteqn{
{\rm F}_{k}(x) =
\frac{1}{(2\pi)^{1+k/2}}
}
\nonumber \\ && \hspace{2ex} \times \,
\sum_{n=1}^{\left[\frac{k-1}{2}\right]} 
\frac{(k\!-\!n)!}{(k\!-\!2n)!}\,{\rm H}_{k-2n}(x)
\int_{-\infty}^{+\infty} dt_1 \, e^{-t_1^2} \, \cdots \,
\int_{-\infty}^{+\infty} dt_{k} \,
\frac{e^{-k t_{k}^2}}{z_{k}^{n}} \, , 
\label{polynomial}
\end{eqnarray}
where the summation index $j$ has been changed as 
\mbox{$m$ $\!-$ $\!j$ $\!=$ $\!n$}. Next we apply the sum 
rules \cite{Prudnikov1}
\begin{equation}
\sum_{j=0}^{\infty}\frac{\Gamma (a + j) z^j}{(2j)! \Gamma (a)} \,
{\rm H}_{2j}(x)
= \frac{1}{(1+z)^a} \, \Phi[a,1/2,x^2 z/(1+z)],
\label{Ma8}
\end{equation}
\begin{equation}
\sum_{j=0}^{\infty}\frac{\Gamma (a + j) z^j}{(2j + 1)! \Gamma (a)} \,
{\rm H}_{2j\!+\! 1}(x)
= \frac{2x}{(1 + z)^a} \, \Phi[a,3/2,x^2 z/(1+z)]
\label{Ma82}
\end{equation}
[$|z|$ $\!<$ $\!1$; $\Phi (a,b,y)$, confluent hypergeometric function]
to the $j$ sums in \mbox{Eqs.~(\ref{Ma72a})} and (\ref{Ma7a}) and obtain 
\begin{eqnarray}
\lefteqn{
K_{2m+1}(x)=
\frac{2x(m+1)!}{(2\pi)^{m+3/2}}
\int_{-\infty}^{+\infty} dt_1 \, e^{-t_1^2} \, \cdots
}
\nonumber \\ && 
\times \, \cdots
\!\int_{-\infty}^{+\infty}
dt_{2m+1} \, e^{-(2m+1)t_{2m+1}^2} \,
\frac{\Phi[m\!+\!2,3/2, z_{2m+1}(1\!+\!z_{2m+1})^{-1}x^2]}
{z_{2m+1}^m(1\!+\!z_{2m+1})^{m+2}} \, ,
\label{kernod}
\end{eqnarray}
\begin{eqnarray}
\lefteqn{
K_{2m}(x)=
\frac{m!}{(2\pi)^{m+1}}
\int_{-\infty}^{+\infty} dt_1 
\, e^{-t_1^2} \, \cdots
}
\nonumber \\ && 
\times \, \cdots 
\int_{-\infty}^{+\infty} dt_{2m} \,
e^{-2mt_{2m}^2}
\,\bigg\{
\frac{\Phi[m\!+\!1,1/2, z_{2m}(1\!+\!z_{2m})^{-1}x^2 ]}
{z_{2m}^m(1\!+\!z_{2m})^{m+1}}-\frac{1}{z_{2m}^m}
\bigg\} .
\label{kernev}
\end{eqnarray}

To write the multidimensional integrals in Eqs.~(\ref{kernod}) and 
(\ref{kernev}) in a more compact form, we change the variables as,
on using generalized spherical coordinates 
\begin{eqnarray}
t_{i} = r \cos \varphi_{i} \prod_{j=1}^{i-1} \sin \varphi_{j}
\quad {\rm if} \quad
i < k,
\quad {\rm and} \quad
t_{k} = r \prod_{j=1}^{k-1} \sin \varphi_{j},
\label{sphercoord}
\end{eqnarray}
and $r$ $\!=$ $\!r_{k}$,
with $0$ $\!\leq$ $\!r$ $\!<$ $\!\infty$,
$0$ $\!\leq$ $\!\varphi_{j}$ $\!\leq$ $\!\pi$ if $j$ $\!<$ $\!k\!-\!1$,
and $0$ $\!\leq$ $\!\varphi_{k-1}$ $\!\leq$ $\!2\pi$. Introducing
the function
\begin{eqnarray}
\lefteqn{
\Omega^{(k)}(r^2) = \int_{0}^{\pi} d\varphi_1 \, 
e^{-t_{1}^{2}} \sin^{k-2}\varphi_1
\, \cdots
\int_{0}^{\pi} d\varphi_{j} \, 
e^{-jt_{j}^{2}} \sin^{k-j-1}\varphi_{j}
\, \cdots
}
\nonumber \\ && \hspace{15ex} \times \,
\cdots 
\int_{0}^{2\pi} d\varphi_{k-1} \, 
e^{-(k-1)t_{k-1}^{2}} 
e^{-kt_{k}^{2}} 
\label{defo}
\end{eqnarray}

and recalling Eq.~(\ref{Ma72aa}) [together with Eq.~(\ref{Ma6a})],
we find that Eqs.~(\ref{kernod}) and (\ref{kernev}) can
be rewritten as given in Eqs.~(\ref{A1}) and (\ref{A2}) in Sec.~\ref{S3}.
Finally, we expand in Eq.~(\ref{polynomial}) $z_{k}^{-n}$ as
\begin{equation}
\frac{1}{z_{k}^{n}} = 
\frac{(-2)^n}{(1-e^{-r_{k}^2})^n} = (-2)^n
\sum_{l=0}^{\infty}\frac{(n\!+\!l\!-\!1)!}{l!(n\!-\!1)!}
\, e^{-lr_{k}^2}
\end{equation}
and perform the Gaussian integrals to obtain F$_{k}(x)$ in the
form of Eq.~(\ref{Ma10}) in Sec.~\ref{S3}.


\section{Derivation of Eqs.~(\protect\ref{omega1}) and 
(\protect\ref{omega2})}
\label{app1}
\setcounter{equation}{0}
\renewcommand{\theequation}{\Alph{section} \arabic{equation}}

In order to write the function $\Omega^{(k)}(z)$ in the form 
of Eq.~(\ref{omega1}) together with Eq.~(\ref{omega2}), we first
rewrite Eq.~(\ref{defo}) as
\begin{eqnarray}
\Omega^{(k)}(r^2) = \int_{0}^{\pi} d\varphi_1 \, 
 \sin^{k-2}\varphi_1
\, \cdots
\int_{0}^{\pi} d\varphi_{i} \, 
\sin^{k-i-1}\varphi_{i}
\, \cdots
\int_{0}^{2\pi} d\varphi_{k-1} \,  
e^{-X_k}, 
\label{defo1}
\end{eqnarray}
where
\begin{eqnarray}
X_k = t_1^2+2t_2^2+\cdots +k t_k^2 \, ,
\end{eqnarray}
$t_{k}$ being given in Eq.~(\ref{sphercoord}). We then expand
the exponential $e^{-X_{k}}$ in a power series, which
implies the power-series expansion of $\Omega^{(k)}(r^2)$,
\begin{eqnarray}
\Omega^{(k)}(r^2) = \sum_{m=0}^{\infty} A^{(k)}_m \, r^{2m}.
\end{eqnarray}
Here, the expansion coefficients $A^{(k)}_m$ are given by
\begin{eqnarray}
A^{(k)}_m=\frac{(-1)^m}{m!}\int_{0}^{\pi} d\varphi_1 \, 
 \sin^{k-2}\varphi_1
\cdots
\int_{0}^{\pi} d\varphi_{i} \, 
\sin^{k-i-1}\varphi_{i}
\cdots
\int_{0}^{2\pi} d\varphi_{k-1} \,Y_k^m ,
\label{multint}
\end{eqnarray}
where $Y_{k}$ $\!=$ $\!X_{k}/r^{2}$ is an $r$-independent angular function,
$Y_{k}$ $\!=$ $Y_{k}(\varphi_{j})$. Particular integrals of the
type given in Eq.~(\ref{multint}) are calculated in \cite{Prudnikov1}.
They can be used to prove, by induction, that for arbitrary
$k$ and $m$ the coefficients $A_{m}^{(k)}$ can be written in the form given 
in Eq.~(\ref{omega2}).


\section{Asymptotics of $K_{k}(x)$}
\label{app2}
\setcounter{equation}{0}
\renewcommand{\theequation}{\Alph{section} \arabic{equation}}

In order to find the asymptotic behaviour of $K_{k}(x)$ for large $x$,
we change the variables in the integrals in Eqs.~(\ref{A1})
and (\ref{A2}) according to \mbox{$\tanh(r^{2}/2)$ $\!=$ $\!\gamma$}
and obtain
\begin{eqnarray}
\label{appt1}
K_{2m+1}(x) = x \int_{0}^{1} d\gamma \, D^{(2m\!+\!1)}(\gamma)
\, \Phi({\textstyle m\!+\!2,\frac{3}{2},-\gamma\,x^2}) 
\end{eqnarray}
and
\begin{eqnarray}
\label{appt2}
K_{2m}(x) =  \int_{0}^{1} d\gamma \, D^{(2m)}(\gamma)
\left[
\Phi({\textstyle m\!+\!1,\frac{1}{2},-\gamma\,x^2})-
(1\!+\!\gamma)^{-(m+1)}
\right] ,
\end{eqnarray}
where
\begin{eqnarray}
\label{appt3}
D^{(2m\!+\!1)}(\gamma) =\frac{2(-1)^m(m+1)!}{(2\pi)^{m+3/2}}
\left[\ln\left(\frac{1+\gamma}{1-\gamma}\right)\right]^{m-1/2}
\nonumber \\ \hspace{5ex} \times \,
\Omega^{(2m+1)}\!\left[ \ln\!\left(\frac{1+\gamma}{1-\gamma}\right)\right]
\frac{(1+\gamma)^{2m+1}}{\gamma^m(1-\gamma)} \, ,
\end{eqnarray}
\begin{eqnarray}
\label{appt4}
D^{(2m)}(\gamma) = \frac{(-1)^m\,m!}{(2\pi)^{m+1}}
\left[\ln\left(\frac{1+\gamma}{1-\gamma}\right)\right]^{m-1}
\nonumber \\ \hspace{5ex} \times \,
\Omega^{(2m)}\!\left[ \ln\left(\frac{1+\gamma}{1-\gamma}\right) \right]
\frac{(1+\gamma)^{2m}}{\gamma^m (1-\gamma)} \, .
\end{eqnarray}

Let us first draw attention to $K_{2m+1}(x)$ in Eq.~(\ref{appt1}). 
For $\gamma$ $\!>$ $\!0$ the function $D^{(2m\!+\!1)}(\gamma)$ 
is finite, and from the asymptotic behaviour of $\Omega ^{(k)}(r^{2})$ 
for large $r$ it follows that $D^{(2m\!+\!1)}(1)$ $\!=$ $\! 0$. From 
the asymptotic behaviour of the confluent hypergeometric function,
\begin{eqnarray}
\Phi(a,c,-t)\approx\frac{\Gamma(c)}{\Gamma(c\!-\!a)}
\,\frac{1}{t^{a}}, \qquad t\to \infty ,
\label{app5}
\end{eqnarray}
it follows that $\Phi({\textstyle m+2,\frac{3}{2},-\gamma\,x^2})$
$\!\to$ $\!0$ if $|x|$ $\!\to$ $\!\infty$ 
except for $\gamma$ $\!=$ $\!0$.
Hence, for \mbox{$|x|$ $\!\gg$ $\!1$} we can approximate the integral in 
Eq.~(\ref{appt1}), replacing $D^{(2m\!+\!1)}(\gamma)$ with its
expansion for $\gamma$ $\!\ll$ $\!1$.
Taking into account that
\begin{eqnarray}
\ln\!\left(\frac{1+\gamma}{1-\gamma}\right) = 2\gamma
\left[ 1 + {\cal O} (\gamma ^{2}) \right] 
\label{app5a}
\end{eqnarray}
and recalling the expansion of $\Omega^{(k)}(z)$, Eqs.~(\ref{omega1}) 
and (\ref{omega2}),
we find that
\begin{eqnarray}
\Omega^{(k)}\!\left[\ln\left(\frac{1+\gamma}{1-\gamma}\right)\right]
= \frac{2\pi^{k/2}}{\Gamma(k/2)} \left[
1 - (k\!+\!1)\gamma + {\cal O} (\gamma ^{2})
\right] .
\label{app6}
\end{eqnarray}
Thus, for $\gamma$ $\!\ll$ $\!1$ the function $D^{(2m\!+\!1)}(\gamma)$,
Eq.~(\ref{appt3}), can be given by
\begin{eqnarray}
\label{appt5}
D^{(2m\!+\!1)}(\gamma) = \frac{(-1)^{m}(m\!+\!1)!}{\pi \Gamma(m+1/2) }
\, \gamma ^{-1/2} \left[ 1 + {\cal O} (\gamma ^{2}) 
\right] ,
\quad \gamma \ll 1 ,
\end{eqnarray}
so that Eq.~(\ref{appt1}) for $x$ $\!\gg$ $\!1$ can be written as, 
on changing the variables as $\gamma ^{1/2} x$ $\!=$ $\!t$, 
\begin{eqnarray}
\lefteqn{
\label{appt6}
K_{2m+1}(x) = \frac{2(-1)^{m}(m\!+\!1)!}{\pi \Gamma( m+1/2 )}
\int_{0}^{x} dt \, \left[ 1 + {\cal O}\!\left( t^{4}/x^{4} \right)
\right]
\Phi({\textstyle m\!+\!2,\frac{3}{2},-t^2}) 
}
\nonumber \\ && \hspace{0ex} 
=  \frac{2(-1)^{m}(m\!+\!1)!}{\pi \Gamma( m+1/2 )}
\left\{ I^{(1)}(0) - I^{(1)}(x) 
+ {\cal O}(x^{-4}) \left[I^{(2)}(0) - I^{(2)}(x)\right] \right\} ,
\end{eqnarray}
where the abbreviating notations
\begin{eqnarray}
\label{appt7}
I^{(1)}(x) = \int_{x}^{\infty} dt \, \Phi({\textstyle m+2,\frac{3}{2},-t^2})
\end{eqnarray} 
and
\begin{eqnarray}
\label{appt8}
I^{(2)}(x) = \int_{x}^{\infty} dt \, t^4 \, 
\Phi({\textstyle m\!+\!2,\frac{3}{2},-t^2})
\end{eqnarray} 
have been introduced.
The integral $I^{(1)}(0)$ can be calculated to be \cite{Erdelyi1}
\begin{eqnarray}
I^{(1)}(0)=\int_{0}^{\infty} dt \, \Phi({\textstyle m+2,\frac{3}{2},-t^2})
= \frac{\pi (2m+1)}{8(m+1)!} \, \Gamma[(2m+1)/2]  
\end{eqnarray}
and it can be shown that $I^{(2)}(0)$ $\!=$ $\! 0$. Further, from
Eq.~(\ref{app5}) it follows that 
$I^{(1)}(x)$ $\!=$ $\!{\cal O}(x^{-2m+1})$
and $I^{(2)}(x)$ $\!=$ $\!{\cal O}(x^{-2m+1})$. For $x<0$ ($|x|$ $\!\gg$ $\!1$)
the calculations are quite analogous, so that 
\begin{eqnarray}
K_{2m+1}(x) =
\textstyle\frac{1}{4} (-1)^m(2m+1) \, {\rm sign}\,(x) 
\left[ 1  + {\cal O}(x^{-2m-3}) \right], \;\; |x| \gg 1.
\label{app14} 
\end{eqnarray}

To find the asymptotic behaviour of $K_{2m}(x)$, we subdivide the 
interval of integration in Eq.~(\ref{appt2}) as
\begin{eqnarray}
\label{appte1}
K_{2m}(x) = I(0,M/x^2) + I(M/x^2,\gamma_{0}) + I(\gamma_{0},1),
\end{eqnarray}
where
\begin{eqnarray}
I(a,b) = \int_{a}^{b} d\gamma \, D^{(2m)}(\gamma) 
\left[
\Phi({\textstyle m\!+\!1,\frac{1}{2},-\gamma\,x^2})-(1+\gamma)^{-m-1}
\right] , 
\label{appte1a}
\end{eqnarray}
and $\gamma_{0}$ $\!\ll$ $\!1$ and $M$ $\!\gg$ $\!1$ such that 
$M/x^{2}$ $\!<$ $\!\gamma_{0}$. 
This reflects the qualitatively different behaviour of the integrand 
in these intervals. In the integrals $I(0,M/x^2)$ and $I(M/x^2,\gamma_{0})$
 the variable
$\gamma$ is small, so that $D^{(2m)}(\gamma)$ can be expanded as,
on using Eqs.~(\ref{app5a}) and (\ref{app6}),
\begin{eqnarray}
\label{appte2}
D^{(2m)}(\gamma) =
(2\pi)^{-1} (-1)^m m \, \gamma ^{-1} \left[ 1 + {\cal O} (\gamma ^{2})
\right], \quad \gamma \ll 1 .
\end{eqnarray}
We also expand $(1\!+\!\gamma)^{-m-1}$ and can rewrite $I(0,M/x^2)$ as
(after changing the variables) 
\begin{eqnarray}
\label{appte3}
I(0,M/x^2) 
= C_{1}(M) + {\cal O}\!\left( M/x^{2} \right) ,
\end{eqnarray}
where
\begin{eqnarray}
\label{appte4}
C_{1}(M) = (2\pi)^{-1} (-1)^{m} m  \int_{0}^{M}
\frac{dt}{t} \left[
\Phi ({\textstyle m\!+\!1,\frac{1}{2},-t}) - 1 
\right] .
\end{eqnarray}
$I(M/x^2,\gamma_{0})$ and $I(\gamma_{0},1)$ can be calculated
integrating the two terms  
in Eq. (\ref{appte1}) separately. Recalling Eq.~(\ref{app5}) and
changing the variables, we find that 
$I(M/x^2,\gamma_{0})$ $\!=$ $\!I^{(a)}$ $\!+$ $\!I^{(b)}$, 
with
\begin{eqnarray}
\label{appte5}
I^{(a)}
& = & (2\pi)^{-1} (-1)^m m 
\int_{M}^{\gamma_{0}x^{2}}
\frac{dt}{t} \left[ 
1 + {\cal O}\!\left( t^{2}/x^{4} \right) 
\right]
\Phi ({\textstyle m\!+\!1,\frac{1}{2},-t}) 
\nonumber \\
& = & {\cal O}\!\left[ (\gamma _{0} x^{2})^{-m-1} \right] + 
C_{2}(M) , 
\end{eqnarray}
where $C_{2}(M)$ $\!=$ $\!{\cal O}(M^{-m-1})$, and [after 
expanding $(1\!+\!\gamma)^{-m-1}$]
\begin{eqnarray}
\label{appte8}
I^{(b)}
& = & (2\pi)^{-1} (-1)^m m 
\int_{M/x^{2}}^{\gamma_{0}} \frac{d\gamma}{\gamma}
\left[ 1 + {\cal O}(\gamma ^{2}) \right]
\left[ 1  + {\cal O}(\gamma )\right] 
\nonumber \\
& = & \pi^{-1} (-1)^{m+1} m \,\ln |x| 
+ C_{3}(M) + C_{4}(\gamma _{0})
+ {\cal O}\!\left( M/x^{2} \right) ,
\end{eqnarray}
where
\begin{eqnarray}
\label{appte9}
&& C_{3}(M) = \pi^{-1}(-1)^{m+1} m  \ln M  ,
\\ && 
\label{appte9a}
C_{4}(\gamma _{0}) = (2\pi)^{-1} (-1)^{m+1} m \left[
\ln \gamma _{0} +  {\cal O}(\gamma _{0} )
\right] .
\end{eqnarray}
Finally, $I(\gamma_{0},1)$ can be written as, on using Eq.~(\ref{app5}),
\begin{eqnarray}
\label{appte10}
I(\gamma_{0},1)
& = & \int_{\gamma_{0}}^{1} d\gamma \, D^{(2m)}(\gamma) 
\left\{ 
{\cal O}\!\left[ (\gamma x^{2})^{-m-1} \right] 
- (1+\gamma)^{-m-1} 
\right\} 
\nonumber \\
& = & C_{5}(\gamma _{0}) +
{\cal O}\!\left( \gamma_{0}^{-m} x^{-2m-2} \right) ,
\end{eqnarray}
where
\begin{eqnarray}
\label{appte11}
C_{5}(\gamma _{0}) = - \int_{\gamma_{0}}^{1} d\gamma \,
D^{(2m)}(\gamma) \, (1+\gamma)^{-m-1}  .
\end{eqnarray}
Substituting in Eq.~(\ref{appte1}) for 
$I(0,M/x^2)$, $I(M/x^2,\gamma_{0})$ and $I(\gamma_{0},1)$ 
the expressions derived above, it can be shown that  
the logarithmic divergences of $C_{4}(\gamma _{0})$ and 
$ C_{5}(\gamma _{0})$ cancel for $\gamma _{0}$ $\!\to$ $\!0$ and 
those of $ C_{3}(M)$ and $C_{1}(M)$ cancel for $M$ $\!\to$ $\!\infty$, 
and hence
\begin{eqnarray}
\label{appte12}
\lefteqn{
K_{2m}(x) = \pi^{-1} (-1)^{m+1} m \, \ln |x|
+ {\cal O}\!\left( M/x^{2} \right) 
}
\nonumber \\ && \hspace{12ex}
+{\cal O}\!\left[ (\gamma_{0}x^{2})^{-m-1} \right]
+ C(\gamma_{0},M), \quad  |x| \gg 1.
\end{eqnarray}
Here, the constant $C(\gamma_{0},M)$ is given by the sum of the constants 
$C_{1}$ -- $C_{5}$. Note that for $\gamma_{0}$ $\!\to$ $\!0$ and $M$
$\!\to$ $\!\infty$ the constant $C(\gamma_{0},M)$ becomes independent on
$\gamma_{0}$ and $M$ [for a determination of this irrelevant constant
the value of the integral (\ref{appte11}) must be known].

\end{appendix}


\vspace{4ex}
\noindent
$^{\ast}\,$Permanent address:
Palack\'{y} University, Faculty of Natural Sciences,
Svobody~26, 77146 Olomouc, Czech Republic

\newpage

\newpage

\newpage

\newpage

\newpage

\newpage

\end{document}